\documentclass[a4paper,fleqn]{cas-dc}

\usepackage[numbers,longnamesfirst]{natbib}

\usepackage[dvipsnames]{xcolor}
\usepackage{paralist}
\usepackage{textcomp}
\usepackage{booktabs}
\usepackage{pifont}
\usepackage{tcolorbox}
\usepackage{multirow}
\usepackage[table]{xcolor}

\usepackage{tabularx}
\usepackage{booktabs}
\usepackage{array}

\usepackage{booktabs}
\usepackage{pifont}

\newcommand{\cmark}{\ding{51}}
\newcommand{\xmark}{\ding{55}}

\newcommand{\toolname}{{\texttt{Trace2ATT\&CK}}}
\newcommand{\attack}{ATT\&CK}

\def\tsc#1{\csdef{#1}{\textsc{\lowercase{#1}}\xspace}}
\tsc{WGM}
\tsc{QE}
\newtheorem{definition}{Definition}
\begin{document}
\let\WriteBookmarks\relax
\def\floatpagepagefraction{1}
\def\textpagefraction{.001}

% Short title
\shorttitle{Mapping Kernel-Level Telemetry to MITRE ATT\&CK}    

% Short author
\shortauthors{Lupinacci et al.}  

% Main title of the paper
\title [mode = title]{A Graph-Based Approach for Mapping Kernel-Level Telemetry to MITRE ATT\&CK}

% First author
\author[1]{Matteo Lupinacci}[orcid=0009-0000-2356-398X]
\cormark[1]
\ead{matteo.lupinacci@unical.it}

% Second author
\author[1]{Francesco Blefari}[orcid=0009-0000-2625-631X]
\ead{francesco.blefari@unical.it}

% Third author
\author[1]{Luigi Arena}[orcid=0009-0008-9844-0229]
\ead{luigi.arena@unical.it}

% Fourth author
\author[1]{Angelo Furfaro}[orcid=0000-0003-2537-8918]
\ead{angelo.furfaro@unical.it}

% Address/affiliation
\affiliation[1]{organization={University of Calabria},
            addressline={Via Pietro Bucci}, 
            city={Rende},
            postcode={87036}, 
            state={CS},
            country={Italy}}

% Corresponding author text
\cortext[1]{Corresponding author}

% For a title note without a number/mark
%\nonumnote{}

% Here goes the abstract
% \begin{abstract}
% Mapping observed system behavior to standardized frameworks like MITRE ATT\&CK is essential for threat-informed defense. Existing automated methods mostly depend on Cyber Threat Intelligence reports, which offer only retrospective accounts of attacks. Low-level telemetry, i.e. kernel-level system calls, instead provides evidence of adversary behavior; yet its volume and complexity have limited its use for automated mapping.
% In this work, we present Trace2ATT\&CK, a framework that collects kernel-level events via eBPF, reconstructs the attacker’s session using a graph-based correlation approach, and derives compact graph representations suitable for LLM-based reasoning. These representations are mapped to the MITRE ATT\&CK framework using both pure LLM prompting and retrieval-augmented generation (RAG) grounded in the ATT\&CK knowledge base, producing ranked technique candidates along with supporting rationales. 
% We evaluate Trace2ATT\&CK on 347 Linux Atomic Red Team tests using locally deployed open-weight LLMs. RAG consistently improves ATT\&CK mapping performance over pure prompting, while provenance graph substantially outperforms raw telemetry. These results show that local inference over graph-based behavioral descriptions can make automated ATT\&CK mapping from kernel-level telemetry operationally viable, without compromising data confidentiality.
% \end{abstract}

\begin{abstract}
Mapping observed system behavior to standardized frameworks such as MITRE \attack{} is essential for threat-informed defense. Existing automated methods mostly depend on Cyber Threat Intelligence reports, which offer only retrospective accounts of attacks. Kernel-level telemetry, i.e., system calls, instead provides first-hand evidence of adversary behavior, but its volume and complexity have limited its use for automated mapping. Its content also makes it unsuitable for cloud-based analysis: command lines carry credentials, events carry user identifiers and network endpoints, and during an incident the telemetry reveals which defenses failed.
In this work, we present \toolname{}, a framework that maps kernel-level telemetry to \attack{} techniques entirely within the defender's infrastructure. \toolname{} collects events via eBPF, reconstructs the attacker's activity as a process-centric provenance graph, and condenses it into a compact execution graph. This graph is classified by locally deployed open-weight LLMs, using either pure prompting or retrieval-augmented generation (RAG) grounded in the \attack{} knowledge base, producing ranked technique candidates with supporting rationales.
We evaluate \toolname{} on 347 Linux Atomic Red Team tests with seven open-weight LLMs. The best local configuration reaches an HR@5 of 66.57\% at the technique level. RAG improves over base prompting in every model and setting, by up to $+15.44$ percentage points, and the execution graph substantially outperforms raw telemetry while never exceeding the LLM context window. Finally, a frontier cloud model used as an upper-bound reference quantifies the cost of local inference at 24--28 percentage points. These results show that \attack{} mapping from kernel-level telemetry is feasible without disclosing sensitive evidence to third-party services.
\end{abstract}

% Use if graphical abstract is present
%\begin{graphicalabstract}
%\includegraphics{}
%\end{graphicalabstract}

% Research highlights
\begin{highlights}
\item Maps kernel-level eBPF telemetry to MITRE ATT\&CK techniques using local LLMs
\item Compact weighted execution graphs make provenance tractable for LLM reasoning
\item Taxonomy grounding and RAG improve HR@5 for every model, up to +15.4 points
\item Execution graphs yield nearly twice the correct mappings of full provenance graphs
\item Technique-level accuracy largely holds even when attacks fail to complete
\end{highlights}

%\nocite{*}

% Keywords
% Each keyword is seperated by \sep

\maketitle

\section{Introduction}
\label{sec:intro}
Modern cyber threats have become increasingly sophisticated, evasive, and rapid in adapting to existing defenses. Attackers employ ever-more-complex tactics, techniques, and procedures (TTPs), making it difficult for security teams to detect and respond to incidents in a timely manner. In this context, the MITRE \attack{} framework~\cite{mitre} has emerged as a standard for categorizing and understanding adversary behavior, providing a structured knowledge base of tactics and techniques observed in real-world attacks. Applying \attack{} effectively requires one fundamental step: mapping observed events to the appropriate TTPs. Identifying the specific techniques used by attackers enables a more consistent approach to threat intelligence, incident alerts, and incident response. However, due to the vast volume of security events that are typically observed in real systems, this mapping is a labor-intensive and error-prone process that is difficult to scale.

The current landscape of \attack{} automated mapping has focused primarily on Cyber Threat Intelligence (CTI) reports. Researchers and practitioners have developed methods to extract tactics and techniques from unstructured CTI documentation using natural language processing and machine learning techniques~\cite{wudali2025rule,cheng2025ctinexus,ruiz2026synt,sigmerge}. These approaches have advanced the field by automating the analysis of threat reports, but share a critical limitation: they operate on second-hand information (i.e., the reports). Meanwhile, the actual events occurring on systems (i.e., application logs, kernel-level syscalls, file modifications, etc.) contain evidence of what attackers do. Although these low-level events are where the reality of an attack unfolds, they remain largely untapped for automatic \attack{} mapping. 

To the best of our knowledge, beyond predefined detection rules (Section~\ref{sec:sub:mapping_approaches}), mapping raw or semi-structured host-level telemetry to MITRE \attack{} remains largely manual.
Bridging this gap is far from straightforward. Systems' logs are massive in volume and complex in structure. A single system generates thousands of events per second; thus, extracting meaningful knowledge from these data is inherently difficult, as it involves identifying which events are relevant to an attack, understanding how individual operations relate to each other, and reconstructing the attacker's sequence of actions from low-level primitives. 

Large Language Models (LLMs) are a natural fit for this reasoning step. However, kernel-level telemetry is precisely the kind of data that should not leave the defender's infrastructure. Command lines carry credentials and tokens, events carry user identifiers and network endpoints, and during an incident the telemetry reveals which hosts are compromised and which defenses failed (Section~\ref{sec:confidentiality}). A practical solution must therefore run with locally deployed, open-weight models, which are typically much smaller than frontier models. Representing the attack compactly and faithfully thus becomes doubly important: the model must reason over the whole attack within a bounded context window and with limited capacity.
To address these challenges, we propose \toolname{}, a framework for the automatic mapping of observed system events to MITRE \attack{} tactics and techniques that:
\begin{inparaenum}[(i)]
    \item collects kernel-level telemetry through an eBPF-based monitor;
    \item reconstructs the attacker's session using a graph-based correlation approach that links actions to their consequences on the system, thus creating a comprehensive provenance graph;
    \item maps the resulting graph to \attack{} techniques using LLM-based systems that run entirely on-premise, optionally grounded in the \attack{} knowledge base through local retrieval-augmented generation.
\end{inparaenum}

\toolname{} offers a different perspective from CTI-based approaches: instead of analyzing what others have reported about attackers, we observe and classify attacker behavior directly from the system itself. The proposed approach is validated using the Linux Atomic Red Team attack scenarios~\cite{atomicredteam}.

We structured our investigation around the following research questions:
\begin{itemize}
    \item \textbf{RQ1}: Can MITRE \attack{} techniques be automatically inferred directly from kernel-level 
    telemetry using locally deployed open-weight LLMs? 
    \item \textbf{RQ2}: How large is the performance gap between open-weight models executed locally and a frontier cloud model, i.e., what is the cost of keeping local inference?
    \item \textbf{RQ3}: Does the choice of graph-based representation affect the quality of \attack{} classification?
\end{itemize}

The main contributions of this paper are as follows:
\begin{itemize}
    \item We present %\href{https://anonymous.4open.science/r/Trace2ATTaCK-C7BF/}{\textcolor{Cerulean}{\underline{\toolname{}}}} 
    \toolname{}, a framework for mapping kernel-level host telemetry to MITRE TTPs, grounding the mapping process directly in observed system behavior rather than relying on post-hoc CTI reports or pre-annotated security rules, and runs entirely within the defender's infrastructure.
    \item We evaluate \toolname{} on 347 Atomic Red Team attack scenarios with seven locally deployed open-weight LLMs under two classification paradigms (pure prompting and RAG). Using a frontier cloud model as an upper-bound reference, we quantify the cost of local inference.
    \item We release a \href{https://anonymous-hf.up.railway.app/a/0jwik3uppjx8/}{\textcolor{Cerulean}{\underline{dataset}}} of the kernel-level logs collected while executing Atomic Red Team scenarios, without any preprocessing, to foster future research in this area. 
\end{itemize}

The remainder of this paper is organized as follows. Section~\ref{sec:back&mot} reviews the background and prior approaches to MITRE \attack{} mapping. Section~\ref{sec:threatmodel} states the threat model and motivates the need for local inference. Section~\ref{sec:methodology} presents the overall framework and methodology, while Section~\ref{sec:dataset} describes the evaluation dataset. In Section~\ref{sec:evaluation} we evaluate \toolname{}, report the experimental results and the ablation study. Section~\ref{sec:related-work} reviews the related works and finally, Section~\ref{sec:conclusion} concludes the paper.

\section{Background and Motivation}
\label{sec:back&mot}

\subsection{Low-level Telemetry}
Telemetry, in the context of computer security, refers to the automated and continuous collection of operational data from systems under observation, enabling real-time monitoring, anomaly detection, and post-incident forensic analysis~\cite{forrest1998intrusion}. Low-level telemetry is the fine-grained variant of this concept, capturing source-adjacent signals such as packet-level, process-level, kernel-level, and device-level events, rather than only aggregated summaries. When telemetry is captured at the kernel-level, through system-call tracing, audit logs, or provenance recording, it provides a privileged view of system activity that is fundamentally harder for adversaries to evade or manipulate than user-space logs~\cite{pasquier2018runtime,bates2015trustworthy}. 
%Kernel-level telemetry captures events such as process executions, file operations, network connections, and inter-process communications. 
In cybersecurity, this granularity is crucial because it provides the observability needed to detect suspicious behavior early, correlate events across heterogeneous assets, and support forensic reconstruction after compromise~\cite{han2020unicorn,cheng2024kairos}.
Given this demonstrated effectiveness in reconstructing sophisticated attack chains, kernel-level telemetry constitutes a reliable foundation for security analytics.

\paragraph{\textbf{eBPF-based telemetry.}} There are several ways to collect kernel-level telemetry. Among them, one of the most prominent and efficient is the extended Berkeley Packet Filter (eBPF)~\cite{ebpf}. eBPF is a technology that enables the safe execution of sandboxed programs directly within the Linux kernel without requiring kernel modifications or module loading~\cite{zhang2024hive}. It has become a critical component in modern Linux systems and is widely adopted by cloud providers to enhance container security, network management, and system observability~\cite{he2023cross}. Although eBPF was initially developed for Linux systems, the advantages it has introduced have led to a current project that aims to introduce eBPF even for Windows systems~\cite{githubGitHubMicrosoftebpfforwindows}.

Compared to traditional kernel audit frameworks (e.g., \texttt{auditd}), eBPF offers three key advantages for security telemetry. 
First, programs are statically verified before execution, ensuring they cannot crash or compromise the kernel, making eBPF a safer alternative to loadable kernel modules \cite{zhang2024hive,lu2024moat}. 
Second, eBPF programs can be attached to a wide range of kernel hook points -- including system call tracepoints, kprobes, and perf events -- at runtime without recompilation or reboot, providing unmatched flexibility in selecting which events to monitor~\cite{he2023cross}. 
Third, because eBPF programs execute in kernel space with Just-In-Time compilation to native instructions, they can filter and aggregate telemetry data before it reaches user space, significantly reducing the performance overhead associated with high-volume event collection~\cite{zhang2024hive}. 

These properties make eBPF particularly well-suited as a low-level telemetry mechanism for cybersecurity, combining the depth of kernel-level observation with the safety, flexibility, and efficiency~\cite{he2023cross,lu2024moat}.

\subsection{The MITRE \attack{} Framework}
\label{sec:sub:mapping_approaches}
MITRE \attack{} (Adversarial Tactics, Techniques, and Common Knowledge) is a curated knowledge base of adversary behavior derived from real-world observations~\cite{mitre,mitredesignphilosofy}. It is organized into matrices for different technology domains (Enterprise, Mobile, and ICS). In this work we use the Enterprise matrix, which covers Linux, Windows, macOS, cloud, and container environments.

The framework describes adversary behavior at increasing levels of detail. \emph{Tactics} capture the adversary's goal, i.e., \emph{why} an action is performed. \emph{Techniques} describe \emph{how} that goal is achieved, and \emph{sub-techniques} refine a technique into more specific variants. Finally, \emph{procedures} are the concrete implementations of a technique observed in the wild, such as a specific sequence of commands. A technique can serve more than one tactic, and closely related techniques often rely on the same low-level operations. As a consequence, attributing an action to a single technique is inherently ambiguous when the action is observed in isolation~\cite{mitredesignphilosofy,mitreAmbiguousTechniques}.

Each technique has a stable identifier, a textual description, and links to the groups, software, mitigations, and detections associated with it. The knowledge base is versioned, updated through periodic releases, and distributed in machine-readable STIX~2.1 format~\cite{mitrekb}. These properties have made \attack{} a common vocabulary for threat intelligence, detection engineering, and adversary emulation~\cite{alsada2024mitre}. 

Tools and frameworks have been developed to assess the efficacy of security measures against the MITRE \attack{} framework. 
Enterprise platforms such as DeTT\&CT~\cite{DeTTECT}, Splunk Enterprise Security~\cite{SplunkEnterpriseSecurity}, and Elastic Security~\cite{ElasticSecurity} follow the same rule-driven paradigm. Their coverage is therefore bounded by the detection rules that have already been written, which in practice cover only a fraction of \attack{} techniques~\cite{CardinalOps2025SIEM}, and they cannot attribute emerging techniques or attack variants that have not yet been modeled.

Mapping observed activity to the correct technique and sub-technique is therefore the step that connects raw evidence to this shared vocabulary, and it is the task addressed by \toolname{}.

\subsection{A Real-World Motivating Example}
Consider the SolarWinds supply-chain compromise~\cite{mitreSolarWindsCompromise}, attributed to APT29 and disclosed in December 2020 after attackers had remained inside victim networks for roughly 14 months. The campaign chained over 70 MITRE \attack{} techniques. Crucially, the \attack{} mapping was produced only months after the initial compromise through manual correlation of reports from FireEye, Microsoft, and CrowdStrike; CISA released its technique table only after public disclosure. In the meantime, victim organizations already had raw telemetry in their logs, such as DNS beaconing, suspicious \texttt{rundll32.exe} invocations, and SAML token anomalies, but lacked the contextual mapping needed to elevate these signals to technique-level alerts.

This case highlights a key limitation of CTI-driven mapping pipelines: classification latency. Mandiant and IBM X-Force reports~\cite{costDetection} show that the median dwell time for advanced threats can range from 100 to over 200 days. 
Our approach targets the step that follows detection: once an alert or a triage process delimits the attack window, \toolname{} maps the observed behavior to \attack{} directly from telemetry, without waiting for post-hoc intelligence.
Beyond reducing latency, this shift could support fast structured alerting, the generation of detection rules, and adversary emulation. Moreover, by attaching a rationale to the proposed mapping, the output can be consumed by SOAR platforms, translated into Sigma rules, and leveraged in purple teaming exercises.

\section{Threat Model}
\label{sec:threatmodel}
Similar to prior research~\cite{depcomm,ocr-apt}, our work considers attackers having access to the target system (locally or remotely) and attempting to maintain a persistent presence by exploiting software vulnerabilities and chaining privilege escalation operations. Our trusted computing base includes the underlying system kernel and the audit framework, along with the offline analysis pipeline code, which is also standard among existing works. 

Any kernel-level attacks that deliberately compromise security auditing systems are beyond the scope of this work. As such, we assume the use of existing system hardening techniques to mitigate any potential audit framework compromise. Moreover, we do not consider hardware-level or side-channel attacks, since their behavior is typically not explicitly captured by kernel-level audit systems. 

We also assume an \emph{honest-but-curious} external AI provider: any third-party inference service lies outside the trusted computing base, so telemetry must not leave the defender's infrastructure.

\subsection{The need for local inference}
\label{sec:confidentiality}
The input of \toolname{} is an unfiltered record of everything that happened on a server during the attack window, which makes it problematic to send to a third-party inference service for several reasons. For example, every \texttt{execve} event carries the full argument vector of the invoked process, together with user identifiers, file paths, and network endpoints. 
%Command lines routinely contain credentials, API tokens, and connection strings.

MITRE catalogues the exposure of sensitive data through process arguments and environment variables as a weakness in its own right (CWE-214~\cite{cwe214}). More broadly, an analysis of public log datasets found that system logs commonly contain personally identifiable information and quasi-identifiers that enable re-identification~\cite{aghili2024privacy}.
 
However, the most critical part is that during incident response, the telemetry exposes the defender's own weaknesses: which hosts are compromised, which privilege-escalation paths succeeded, and which defenses failed. %Sending it to a remote provider extends the attack surface to that provider's infrastructure, where prompts and logs may be retained and become a target in their own right.
This risk is not hypothetical. Researchers found a publicly exposed database of an LLM provider containing user chat histories and API secrets~\cite{wiz_deepseek2025}. Data submitted to a model can also resurface through memorization if it is later used for training~\cite{carlini2021extracting}. Users also routinely disclose sensitive information to hosted LLMs, often unintentionally~\cite{mireshghallah2024trustnobot}.
 
These considerations motivate the \emph{local execution} design goal of \toolname{}.

\section{\toolname{}: Methodology and Construction}
\label{sec:methodology}
\begin{figure*}[ht]
    \centering
    \includegraphics[width=\linewidth]{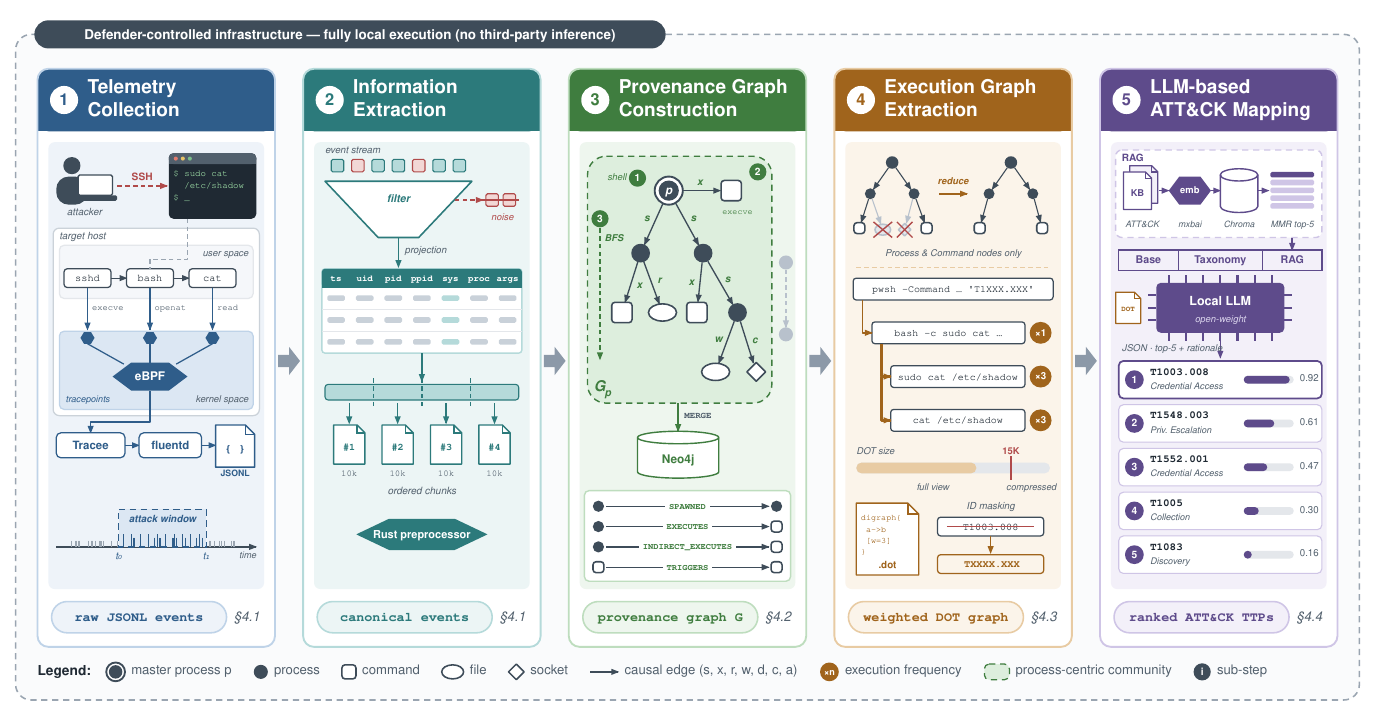}
    \caption{Overview of the proposed methodology for mapping eBPF-based kernel telemetry to MITRE ATT\&CK.}
    \label{fig:methodology}
\end{figure*}
%

% \revml{Tempi del framework:}
% \begin{verbatim}
% [*] [301/417] T1552-003-2 (atomicRedTeam, root)
% [*] Building environment...
% [DONE] Environment rebuilt  (16s)
% [*] Executing attack...
% [*] Waiting 15s for log buffering...
% [*] Starting Trace2ATT&CK for T1552-003-2...
% [DONE] Preprocessing logs   (1.9s)
% [DONE] Database build       (4.4s)
% [DONE] Provenance graph     (1.1s)
% [DONE] Execution graph      (0.9s)
% [DONE] Trace2ATT&CK completed -> outputs/T1552-003-2/
% \end{verbatim}

This section presents \toolname{}, a framework for automatically enriching event sequences with MITRE \attack{} TTPs and transparent rationales. It is worth noting that we use the v19 \attack{} matrix, released in April 2026.

As shown in Figure~\ref{fig:methodology}, the pipeline comprises five phases: \textit{log collection, information extraction} through log preprocessing, \textit{provenance graph construction, execution graph extraction} through graph reduction, \textit{TTPs extraction} across the entire attack lifecycle and explainability using LLM-based systems.

Three design goals guide the system:
\begin{itemize}
    \item \textbf{Context sensitivity}. The system should recognize that many meaningful MITRE \attack{} inferences depend on event sequences, causality, and process relationships rather than isolated activities. Indeed, a command's maliciousness often depends on its contextual TTPs. For example, the command \texttt{bash -c "rm -rf /tmp/logs/*"} may indicate routine cleanup in one context but forensic evidence deletion in another~\cite{sigmerge}.
    \item \textbf{Local execution}. Kernel-level telemetry contains credentials, personal data, and evidence of an ongoing compromise (Section~\ref{sec:confidentiality}). The full pipeline (LLM inference, embeddings, and vector store) must therefore run on infrastructure controlled by the defender, without disclosing telemetry to third-party services, supporting on-premise and privacy-sensitive use cases.
    \item \textbf{Interpretability and Explainability}. The system should not only predict techniques but also explain why the evidence supports those techniques.
\end{itemize}

We focus on a predefined temporal window in which the attack is known to have occurred. Accordingly, our objective is not anomaly detection but contextual explainability, enrichment, and reconstruction of attack behavior. While anomaly detection may be used as a preceding step to identify such a window, it is outside the scope of this work. In practice, any anomaly detector or incident triage mechanism could serve this purpose, but our method assumes that the attack interval has already been delimited and therefore concentrates on fine-grained semantic enrichment.

\subsection{Telemetry Collection and Preprocessing}
\label{subsec:telemetry}

\begin{table}[ht]
\centering
\caption{Mapping between the collected Tracee events and the provenance relations of Section~\ref{sec:graph-model}.}
\label{tab:events}
\small
\setlength{\tabcolsep}{4pt}
\begin{tabular}{@{}l l l@{}}
\toprule
\textbf{Tracee event(s)} & \textbf{Relation} & \textbf{Edge} \\
\midrule
event context$^{\dagger}$ & $R_s$ (spawn) & $\mathcal{P}\to\mathcal{P}$ \\
\texttt{execve} & $R_x$ (execute) & $\mathcal{P}\to\mathcal{C}$ \\
\texttt{fs} set$^{\ddagger}$ & $R_r$ (read) & $\mathcal{P}\to\mathcal{F}$ \\
\texttt{fs} set$^{\ddagger}$ & $R_w$ (write) & $\mathcal{P}\to\mathcal{F}$ \\
\texttt{fs} set$^{\ddagger}$ & $R_d$ (delete) & $\mathcal{P}\to\mathcal{F}$ \\
\texttt{security\_socket\_connect} & $R_c$ (connect) & $\mathcal{P}\to\mathcal{N}$ \\
\texttt{security\_socket\_accept}  & $R_a$ (accept) & $\mathcal{P}\to\mathcal{N}$ \\
\bottomrule
\end{tabular}

\vspace{3pt}
\parbox{\columnwidth}{\footnotesize
$^{\dagger}$Every Tracee event carries \texttt{processEntityId} and \texttt{parentEntityId} in its context, so parent--child edges are derived without a dedicated \texttt{fork}/\texttt{clone} event.\\
$^{\ddagger}$The \texttt{fs} set groups the file-system syscalls and LSM hooks traced by Tracee. File relations are derived from the events that carry a path. Open events (\texttt{security\_file\_open}, which covers \texttt{open}, \texttt{openat} and \texttt{openat2}) are mapped to $R_r$ or $R_w$ according to the access mode in their flags (\texttt{O\_RDONLY} vs.\ \texttt{O\_WRONLY}/\texttt{O\_RDWR}); \texttt{creat}, \texttt{truncate} and \texttt{rename*} are mapped to $R_w$; \texttt{unlink*}, \texttt{rmdir} and \texttt{security\_inode\_unlink} to $R_d$. Since opening a file for writing does not imply that data was written, $R_w$ captures write intent. Events without a path (e.g., \texttt{read}/\texttt{write} on file descriptors, \texttt{stat}, \texttt{poll}) are discarded during filtering.}
\end{table}

\toolname{} relies on Tracee~\cite{tracee}, an eBPF-based runtime monitoring tool, to collect kernel-level telemetry from the target system. Tracee captures system calls and kernel events, writing them as structured JSONL records enriched with process metadata. Table~\ref{tab:events} lists the events we capture using Tracee and the provenance relation of Section~\ref{sec:graph-model} that each one populates.

The raw output is processed by a custom preprocessor that performs two operations:
\begin{enumerate}
    \item \textit{Event filtering}. We discard events generated by the monitoring pipeline itself (Tracee and the log forwarder).
    \item \textit{Fixed schema creation}. Each remaining event is mapped onto a fixed schema of fields:
    \begin{inparaenum}[(i)]
        \item timestamp, 
        \item userId, 
        \item entityId, 
        \item parentEntityId, 
        \item syscall, 
        \item processName, 
        \item args;
    \end{inparaenum}
\end{enumerate}
We key processes on Tracee's entity identifiers rather than PIDs, since PIDs are recycled during long sessions, whereas entity identifiers are unique for the lifetime of the trace.

It should be noted that, although we rely on Tracee for kernel-level log collection, this methodological step can be performed using any eBPF program. This is achieved by extracting the set of features described above from the raw log; examples include other eBPF-based tools such as Tetragon~\cite{tetragon} and Falco~\cite{falco}.

% After our preprocessing, the final lines of logs analyzed are 21.823.051 amounting to 9.7 GB. On average, each attack scenario in our dataset is described by more than 62K lines of preprocessed kernel-level telemetry.

\subsection{Graph-based Attack Reconstruction}
\label{sec:graph-model}
While raw log entries may not portray the intrinsic characteristics of a complex cyber attack, Provenance Graphs (PG) preserve the causal relationships between system entities in a unified directed graph essential for reconstructing attack scenarios~\cite{han2018provenance,cheng2024kairos}. This graph-based approach has become \textit{the de facto standard} for detecting and investigating Advanced Persistent Threats (APTs)~\cite{yang2023prographer}. For these reasons, we use PG as a formal model for the attack reconstruction phase, in which nodes represent system entities and edges represent their interactions. 

\toolname{} constructs two graph representations of the occurred attack: a full provenance graph and an execution graph. The reasons behind this choice are detailed in the following paragraph.

\subsubsection{Graph Model}
Let us consider the following sets of system entities:
\begin{itemize}
    \item \(\mathcal{P}\) be the set of \textit{processes}, where each process is an execution context identified by a unique entity identifier and characterized by attributes such as a process name and a user identifier.
    \item \(\mathcal{C}\) be the set of \textit{commands}, where each command represents an operation correlated with a process execution.
    \item \(\mathcal{F}\) be the set of \textit{files}, where each file is a file-system object identified by a path.
    \item $\mathcal{N}$ be the set of \emph{network endpoints}, where each endpoint is identified by an address and a port.
\end{itemize}

Furthermore, let \(T = \{s, x, r, w, d, c, a\}\) be the set of event types, representing the following actions: spawn (\(s\)), execute (\(x\)), read (\(r\)), write (\(w\)), delete (\(d\)), connect (\(c\)), and accept (\(a\)).
For each event type \(t \in T\), we define a causal relationship \(R_t\) as follows:
\begin{itemize}
    \item \(R_s \subseteq \mathcal{P} \times \mathcal{P}\): a process creates another process (establishing a parent-child relationship).
    \item \(R_c, R_a \subseteq \mathcal{P} \times \mathcal{N}\): a process initiates a connection to a remote endpoint, or accepts a connection on a local endpoint, respectively.
    \item \(R_x \subseteq \mathcal{P} \times \mathcal{C}\): a process executes a command.
    \item \(R_r, R_w, R_d \subseteq \mathcal{P} \times \mathcal{F}\): a process accesses a file to read, modify, or delete it, respectively.
\end{itemize}

Our proposed PG is the directed graph $G = (V, E)$ where $V \subseteq \mathcal{P} \cup \mathcal{C} \cup \mathcal{F} \cup \mathcal{N}$
is the set of nodes, and
\[ E =  \bigcup_{t \in T} R_t  \cap \left( V \times V\right )    \]
is the set of edges.
By definition, each edge corresponds to a specific causal relationship between two system entities.
\\\\
\noindent 
We introduce the notion of a \emph{process-centric community} as the analytical unit for attack reconstruction, according to previous research~\cite{depcomm}. 

\begin{definition}[Process-Centric Community]
A process-centric community rooted at a master process $p \in \mathcal{P}$ is the induced subgraph $G_p = (V_p, E_p)$, where $V_p$ contains $p$ and all entities reachable from $p$ via causal edges in $E$, and $E_p$ contains all edges among entities in $V_p$.
\end{definition}

This concept captures the observation that system activity during an attack is not uniformly distributed across the provenance graph, but rather concentrates around a small number of \emph{master processes} that drive the malicious execution. Such a self-contained subgraph represents a coherent unit of adversarial behavior. 

\subsubsection{Full Provenance Graph Construction}
\toolname{} scans the preprocessed logs in chronological order to locate the \texttt{execve} of a recognized shell (e.g., \texttt{bash}, \texttt{sh}) that is not a probe artifact. Following the assumptions given in Section~\ref{sec:methodology}, within the predefined temporal window, this shell is the entry point of the adversarial session. The identified shell serves as the master process $p$ of the community.
%; then, the subsequent phases build the induced subgraph $G_p$ by collecting all entities causally reachable from~$p$.

Then, a key step is to correlate the actions directly issued by the attacker with the consequences they produce in the system. For the attack reconstruction, this is important because a single command may generate several events, all of which have an impact on the system. To retrieve the commands issued by the attacker, \toolname{} relies on 
%a command-collection mechanism based on 
shell logging and \texttt{execve} system calls analysis. %with fallback. 
%Commands can be obtained through shell history files such as \texttt{bash\_history} for Bash and \texttt{zsh\_history} for Zsh, through accounting tools such as \texttt{psacct} on Red Hat-based systems and \texttt{acct} on Debian-based systems, or through a custom trap-based logger.
More in depth, for each directly issued command, \toolname{} identifies the corresponding \texttt{execve} event by matching the command timestamp to the nearest \texttt{execve} occurrence within a bounded lookahead window and its arguments. To preserve command ordering and prevent many-to-one matches, the search position is advanced monotonically after each successful association. 
If shell logging is unavailable (for any reason) or fails to capture a command, the module falls back to directly reconstructing the command line from 
%any remaining, unmatched 
\texttt{execve} events, ensuring that no observed process execution is silently dropped from this step, even in the absence of shell-level logging. 

Finally, \toolname{} reconstructs the full descendant process tree through a breadth-first traversal (BFS) over spawn edges ($R_s$) from $p$, which instantiates $V_p$ of Definition 1: each reached process contributes its reconstructed command line and its complete local event stream (execute, file and network edges), but reachability is not propagated through file or socket entities, which avoids dependency explosion.
%
% Finally, \toolname{} reconstructs the full descendant process tree through a breadth-first traversal (BFS) over spawn edges ($R_s$) relations, and attaches to each discovered subprocess both its reconstructed command line and its complete local event stream. 
% In practice, $V_p$ is instantiated by a BFS over spawn edges ($R_s$) from $p$; each reached process contributes its execute, file and network edges.
%This helps to derive for each activity it's triggered consequences on the host.
This step finalizes the creation of the process-centric community by creating the subgraph $G_p$, that is a provenance graph.

\subsubsection{Execution Graph Extraction}

\begin{figure*}[ht]
\centering
\includegraphics[width=.75\linewidth]{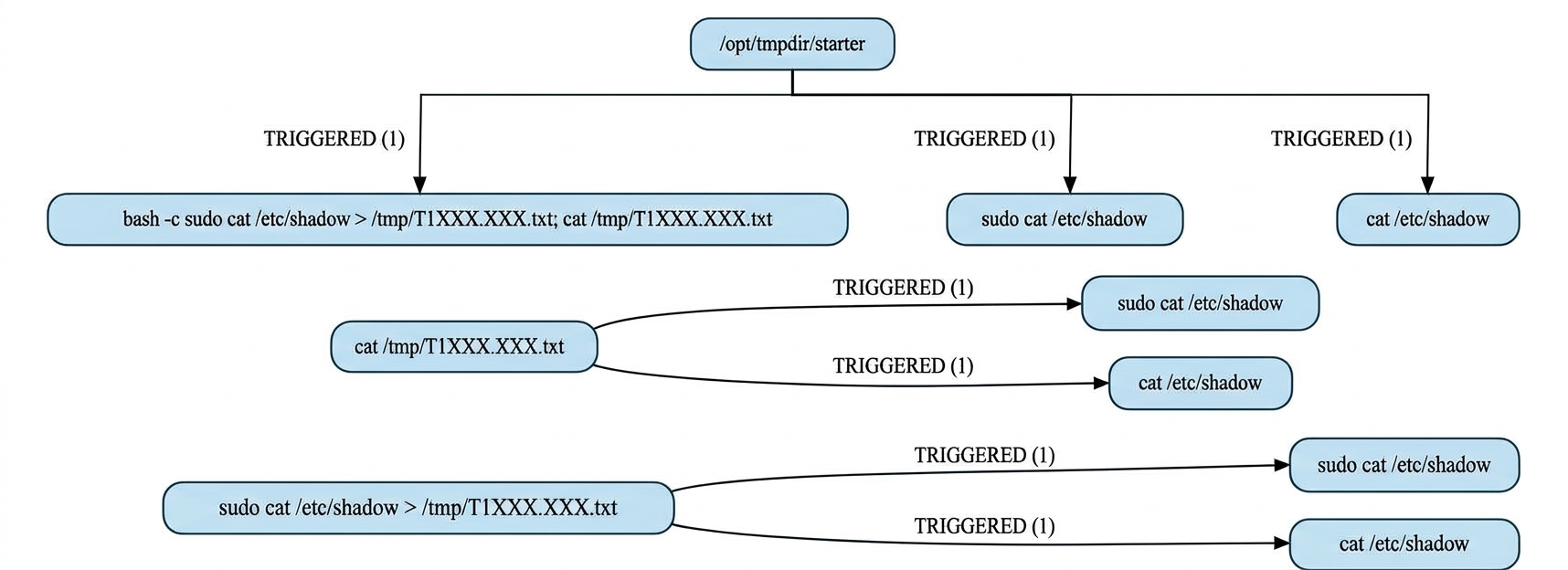}
\caption{Weighted Execution Graph Extracted for T1003.008 (OS Credential Dumping: /etc/passwd and /etc/shadow)}
\label{fig:cmd_graph}
\end{figure*}

After the creation of the full PG, we apply a graph reduction strategy to optimize query performance and maintain a tractable input size for LLM reasoning while preserving the complete provenance information for potential use in other analysis tasks.

Specifically, we apply a PG reduction strategy tailored for the task of MITRE \attack{} mapping by restricting the node set to \texttt{Process} and \texttt{Command} entities, thus obtaining an execution graph (EG). A representative example of the extracted execution graph is shown in Figure~\ref{fig:cmd_graph}. 

This choice is motivated by the nature of the task itself: \attack{} techniques are defined at the level of adversary \emph{procedures} (i.e., the concrete actions performed on the system) and these actions are primarily captured by process executions and their associated command-line arguments. File paths, network endpoints, and other system objects are then embedded within command arguments and can therefore be inferred contextually by the LLM without explicit graph nodes. 

We call this representation an execution graph rather than a command graph, because its nodes are not limited to the commands typed by the attacker. In many real intrusions, and in our evaluation, the adversary issues a single command, such as a launcher script or a one-liner that downloads a script and pipes it into a shell, and every subsequent step is executed by its descendant processes. These steps never appear in the shell history: they are visible only as execve events in the kernel telemetry, from which \toolname{} reconstructs them.

The relations between nodes in the EG are defined as follows:
\begin{itemize}
    \item \texttt{SPAWNED}: parent-to-child process relationships;
    \item \texttt{EXECUTES}: direct command execution by a process;
    \item \texttt{INDIRECT\_EXECUTES}: sub-process command execution;
    \item \texttt{TRIGGERS}: parent-command-to-child-command causal links.
\end{itemize}

Restricting the node set directly addresses the well-documented \emph{dependency explosion} problem in provenance analysis: large, fully expanded provenance graphs are impractical to process within the finite context window of large language models, as injecting comprehensive event histories incurs prohibitive token overhead~\cite{trackagent}. Other research in the field further supports this design: PROV-LLM~\cite{provllm} demonstrates that decomposing large provenance graphs into process-centered subgraphs yields superior detection accuracy and precision compared to full-graph approaches. More evidence about this choice is illustrated in Section~\ref{sec:ablation}.

\subsection{Execution Graph Encoding for LLMs}
\label{subsec:execution-graph}

The process of encoding graph as text is a necessary step for turning graph-structured information into a sequence for reasoning by LLMs.
\toolname{} analyzes the entire execution graph and extracts a \textit{weighted execution graph} in which repeated executions of the same execution pattern are aggregated and represented through a weight proportional to their execution frequency. Weight highlights a useful property for \attack{} mapping: it is not just which commands appear that matters, but also how often they are repeated. Frequency helps distinguish isolated actions from stronger procedural patterns. \toolname{} derives two weighted representations.

\paragraph{\textbf{Full weighted execution graph.}}
The first representation preserves the original execution instances observed during the attack session. Each \texttt{EXECUTES} edge yields a node containing the executing process (entity ID, name, user ID), the command (full command line, timestamp), its privilege-escalation status, and the list of triggered sub-commands retrieved through \texttt{TRIGGERS} edges. In addition, each execution node is annotated with the number of times the same command line has been executed in the session, while repeated triggered sub-commands are aggregated and labeled with their corresponding occurrence count. 

\paragraph{\textbf{Compressed weighted execution graph.}}
The second representation reduces redundancy by normalizing commands (i.e., extracting the base binary and option flags), and merging identical \texttt{command+option} patterns into a single normalized view. As in the raw version, the resulting graph is weighted. This compressed weighted view is especially useful for attacks characterized by repetitive behavior, such as iterative reconnaissance or repeated payload invocation.

%Both representations are serialized as DOT graph files (via \texttt{pydot}). 
When the full weighted graph exceeds 15,000 characters, the system automatically falls back to the compressed representation.

\subsection{TTPs Extraction}
The final step of \toolname{} consists of mapping the observed attack flow to MITRE \attack{} TTPs. 
We adopt LLM-based approaches because these models have demonstrated state-of-the-art performance in processing and reasoning over structured and semi-structured textual data, including security-related graph representations~\cite{lekssays2025llmxcpg}. 
We employ two complementary LLM paradigms: pure LLM prompting and RAG to evaluate which configuration best leverages the \attack{} knowledge base for accurate classification. 
Within the pure-prompting paradigm we consider two prompt variants: a \emph{base} prompt, and a \emph{taxonomy-grounded} prompt that additionally injects the complete mapping from technique IDs to their canonical names (motivated in Section~\ref{sec:results}).
RAG operates by dynamically retrieving relevant \attack{} technique descriptions and providing them as factual context to the LLM, which has been shown to reduce hallucination without requiring fine-tuning. 

For each input graph, the system outputs a ranked list of top-$k$ candidate classifications, where each candidate includes the \attack{} technique identifier, the associated tactic, a confidence score in the range $[0,1]$, and a natural-language explanation justifying the mapping. The inclusion of explanations serves a dual purpose: it increases the explainability of the classification and enables human analysts to validate the LLM's reasoning, which is critical for trustworthy deployment in security operations.

The decision to output a ranked list of  top-$k$ candidate techniques, rather than a single best-match prediction, is motivated by the inherent ambiguity of \attack{} technique attribution. As previously observed in~\cite{lade}, distinct techniques can exhibit highly similar behavioral characteristics at the system level; for instance, T1105 (Ingress Tool Transfer) and T1048 (Exfiltration Over Alternative Protocol) frequently rely on comparable network utilities, making a single-prediction approach prone to false negatives when the model's confidence is distributed among semantically related candidates. A ranked top-$k$ output mitigates this risk by preserving the most plausible alternatives, enabling downstream evaluation metrics and providing security analysts with a broader set of hypotheses to validate.

\paragraph{\textbf{Prompt design.}}
As prompting strategy, \toolname{} adopts a zero-shot in context learning with expert instruction. Indeed, the system prompt instructs the LLM to act as a cyber threat intelligence analyst and to output a JSON array of the top-$K$ candidate techniques ranked by confidence (see Appendix~\ref{sec:appendix_c}).
Each candidate includes the tactic name, the technique with sub-technique  if any (e.g., \texttt{T1003.007}), a confidence score in $(0,1)$, and a natural-language rationale for the proposed mapping.
The prompt enforces two critical constraints:
\begin{inparaenum}[(i)]
    \item sub-techniques must be used whenever the attack evidence warrants them;
    \item rationales must reference the \emph{entire command flow}, not isolated commands.
\end{inparaenum}

\paragraph{\textbf{Retrieval-augmented generation.}}
The RAG pipeline ingests the MITRE \attack{} Enterprise knowledge base~\cite{mitrekb} into a local Chroma vector store.
Technique descriptions are split into 800-word chunks and indexed using the \\ \texttt{mxbai-embed-large-v1}~\cite{emb2024mxbai,li2023angle} embedding model, widely used in AI for semantic search, clustering, classification, and RAG.
At inference time, the DOT graph is embedded, and the top-5 most relevant technique chunks are retrieved via Maximal Marginal Relevance (MMR) search ($\lambda_{\text{MMR}} = 0.5$, fetch\_k = 20).
These chunks are prepended as factual context to the user prompt, grounding the LLM's reasoning in the official \attack{} taxonomy without requiring the full knowledge base to fit in the context window.

\section{Dataset}
\label{sec:dataset}
%\revml{Dire che noi usiamo dataset aggiornati, non come fanno altri tool SOTA tipo Sigmerge che usa dataset vecchi le cui tecniche non esistono più in MITRE \attack{} [T1139, T1145, T1146, T1184, T1168] (i.e., il loro dataset 'metta': https://github.com/uber-common/metta)}

For evaluation, we rely on Atomic Red Team~\cite{atomicredteam}, an open-source library of MITRE ATT\&CK framework~\cite{mitre} tests. The dataset is organized as a structured repository where each test -- referred to as an "atomic test" -- resides in a dedicated directory (e.g., T1055 for Process Injection). 
Each atomic test corresponds to a specific \attack{} technique, providing a pre-defined labeling scheme that eliminates the need for manual annotation. It is worth noting that Atomic Red Team is not a classification tool: its purpose is to simulate real-world attack behaviors in a controlled environment, not to identify or label techniques from observed activity. The ground truth is therefore an intrinsic property of the dataset itself, aligned with a taxonomy widely recognized in both academia and industry~\cite{alsada2024mitre}.

% Each atomic test corresponds to a specific \attack{} technique, providing a pre-defined labeling scheme that eliminates the need for manual annotation and ensures that the ground truth is aligned with a taxonomy widely recognized in both academia and industry~\cite{alsada2024mitre}.
%Each atomic test contains a YAML definition file specifying the test metadata, supported platforms (Linux, macOS, Windows), execution commands, optional prerequisites, input arguments, and cleanup commands.
Every test performs a sequence of actions that can vary from two actions up to thirty actions depending on the scenario. We therefore treat one atomic test as one labeled attack \emph{session} with a single ground-truth technique, that may result in several commands in the telemetry sense.

To execute the atomic tests in a controlled and reproducible manner, it is possible to manually execute the commands specified in the YAML file or use the provided \texttt{Invoke-AtomicRedTeam}~\cite{invokeatomic}, a PowerShell-based execution framework that automates the invocation of tests. 

Leveraging the \texttt{Invoke-AtomicRedTeam} module, the command used to execute a test may explicitly embed the ground-truth \attack{} identifier. If left unchanged, this would make the evaluation unfair, as the LLM could recover the correct label directly from a single command rather than from the reconstructed behavior. Therefore, we sanitize command lines before inference by obfuscating all \attack{} technique and sub-technique identifiers with placeholder tokens (e.g., \texttt{T1059.003} $\rightarrow$ \texttt{TXXXX.XXX}) along with path arguments that could unfairly help the LLM. 

The resulting dataset, obtained by executing all attacks in the evaluation set and collecting logs via Tracee as the eBPF program, consists of 21,950,327 lines of raw logs, amounting to 29.8 GB. On average, each attack scenario is described by more than 62K lines of raw kernel-level telemetry.

\paragraph{\textbf{Data contamination.}} Atomic Red Team is publicly available, so its tests may appear in the pre-training data of the evaluated models. Two factors limit this risk. First, technique and sub-technique identifiers are masked in the command lines, so the label cannot be read from the input. Second, the ground truth follows ATT\&CK v19 (released on 28 April 2026) and the Atomic Red Team release migrated to v19 on 2 May 2026, whereas the training-data cutoff of every evaluated model, including the frontier reference, precedes both dates (Table~\ref{tab:models}). Labels introduced or changed in v19 therefore cannot have been memorized. Contamination cannot be excluded for tests whose commands and identifiers are unchanged from earlier releases.

\section{Evaluation}
\label{sec:evaluation}

\begin{table*}[ht]
\centering
\scriptsize
\caption{LLMs evaluated in our experiments.}
\label{tab:models}
\begin{tabular}{lcccccc}
\toprule
\textbf{Model} & \textbf{Date} & \textbf{Knowledge Cutoff}  & \textbf{Params} & \textbf{Quantization} & \textbf{Native Context} & \textbf{Configured Context} \\
\midrule
    \href{https://huggingface.co/lmstudio-community/gpt-oss-20b-GGUF}{gpt-oss-20b~\cite{agarwal2025gpt}} & 
    August 2025 & 
    June 2024 & 
    20B & 
    GGUF MXFP4 &
    128K &
    32K \\
    
    \href{https://huggingface.co/lmstudio-community/gpt-oss-120b-GGUF}{gpt-oss-120b~\cite{agarwal2025gpt}} & 
    August 2025 & 
    June 2024 &
    120B & 
    GGUF MXFP4 & 
    128K &
    32K \\
    
    \href{https://huggingface.co/unsloth/gemma-4-31B-it}{gemma-4-31b-it~\cite{gemmateam2026gemma4technicalreport}} &
    April 2026 & 
    January 2025 & 
    31B & 
    GGUF Q4\_K\_S & 
    256K &
    32K \\
    
    \href{https://huggingface.co/unsloth/Llama-3.3-70B-Instruct-GGUF}{llama-3.3-70b-instruct~\cite{huggingfaceUnslothLlama3370BInstructGGUFHugging}} & 
    December 2024 & 
    December 2023 & 
    70B & 
    GGUF Q4\_K\_S & 
    128K &
    32K \\

    \href{https://huggingface.co/lmstudio-community/DeepSeek-R1-Distill-Qwen-32B-GGUF}{deepseek-r1-distill-qwen-32b~\cite{huggingfaceLmstudiocommunityDeepSeekR1DistillQwen32BGGUFHugging}} & 
    January 2025 & 
    July 2024 & 
    32B & 
    GGUF Q8\_0 & 
    128K &
    32K \\

    \href{https://huggingface.co/fdtn-ai/Foundation-Sec-8B-Reasoning-Q8_0-GGUF}{foundation-sec-8b-reasoning~\cite{kassianik2025llama}} & 
    January 2026 & 
    April 2025 & 
    8B & 
    GGUF Q8\_0 & 
    32K &
    32K \\

    \href{https://huggingface.co/Qwen/Qwen3.5-9B}{qwen3.5-9b~\cite{qwen3.5}} & 
    February 2026 & 
    January 2026 & 
    9B & 
    GGUF Q8\_0 & 
    262K &
    32K \\

    \href{https://www.anthropic.com/news/claude-sonnet-5}{Claude-Sonnet-5~\cite{claude-sonnet-5}} & 
    June 2026 & 
    January 2026 & 
    N/A & 
    N/A & 
    1M &
    N/A \\
\bottomrule
\end{tabular}
\end{table*} 

We assume a threat scenario in which the attacker has remote access to the target system, for instance through an interactive SSH session, and issues commands from a remote shell. 
An attacker can drive a session in two ways: interactively, typing each step in the remote shell, or through a single invocation of a script or wrapper that performs all steps on its own. Our evaluation adopts the second, harder setting. Every Atomic Red Team test is launched through one command, so shell-level logging observes a single opaque invocation, while the entire procedure, including steps fetched at runtime and piped into an interpreter (e.g., \texttt{curl ... | bash}), is reconstructed exclusively from kernel-level \texttt{execve} events. On average, a single issued command expands into 24 process executions per scenario.

%These commands may correspond either to the actual attack sequence or to a higher-level script, wrapper, or utility that internally triggers the malicious actions. 

The preprocessor is written in Rust and distributed as a cross-platform binary (Linux x86\_64/ARM64, macOS x86\_64/ARM64). 
%After our preprocessing, the final lines of logs analyzed are 21.823.051 amounting to 9.7 GB. 
The PG modules and TTPs extraction using LLMs phase are written in Python. The PG creation is supported by Neo4j, and for execution graph extraction, we create custom Cypher query.

We test \toolname{} using seven local LLMs, listed in Table~\ref{tab:models}, on the Linux-specific subset of test from the Atomic Red Team dataset. 
All metrics reported in this section are computed over the $N_v$ valid mappings remaining after excluding cases in which the LLM output was not a well-formed JSON object, which we treat as unparsable predictions. More details of invalid output per model are shown in Table~\ref{tab:invalid_json_llm_rag}.
Moreover, a subset of tests fails at execution time. Prerequisites were installed with Invoke-AtomicTest -GetPrereqs before each test; the remaining failures stem from binaries referenced but not distributed by the Atomic Red Team repository 
or from insufficient privileges of the executing user.
%
%Moreover, since a subset of tests fails at execution time due to missing dependencies in the Atomic Red Team repository or due to missing permissions of the user executing the attack, we 
%
We therefore report metrics separately for tests successfully concluded (i.e., the test returned status code is 0, $N_s = 166$) and for failed tests ($N_f = 181$). 
Along with results on technique identification across the entire dataset, we report results separately for the subset of tests whose ground truth specifies a sub-technique ($N_{sub} = 214$) by looking for the exact match (e.g., T1003.007). 

\subsection{Evaluation Metrics}
\label{sec:appendix_a}
We report multiple ranking-based metrics~\cite{lade} in order to capture both correctness and placement quality. 

Let $N = N_v$ denote the total number of test cases retained for evaluation, and let $\text{rank}_i \in \{1, \dots, K\}$ denote the position of the correct label in the ranked list of predictions for test case $i$ (with $\text{rank}_i = \infty$ if the correct label does not appear within the top-$K$ candidates).
% We define an indicator function 
%\begin{equation}
%I(\text{rank}_i \leq K) =
%\begin{cases}
%%1 & \shortstack[l]{\text{if the correct label appears within}\\ \text{the top-}K\text{ predictions}} \\
%1 & \parbox{120pt}{if the correct label appears within\\ the top-K predictions} \\
%0 & \text{otherwise.}\\
%
%\end{cases}
%\end{equation}

\paragraph{Hit Rate (HR@$K$).} HR@$K$ measures the percentage of test cases for which the correct label appears anywhere within the top-$K$ ranked predictions, regardless of its exact position:
\begin{equation}
\text{HR@}K = \frac{100}{N} \sum_{i=1}^{N} I(\text{rank}_i \leq K)
\end{equation}

where $I(\text{rank}_i \leq K)$ is an indicator function that evaluates to $1$ if the ground-truth label appears in the top-K predictions and $0$ otherwise.
\paragraph{Mean Reciprocal Rank (MRR@$K$).} MRR@$K$ rewards predictions in which the correct label is ranked closer to the top, assigning a reciprocal-rank score of zero whenever the correct label falls outside the top-$K$:

\begin{equation}
\text{MRR@}K = \frac{1}{N} \sum_{i=1}^{N}
\begin{cases}
\dfrac{1}{\text{rank}_i} & \text{if } \text{rank}_i \leq K \\[6pt]
0 & \text{otherwise}
\end{cases}
\end{equation}

\paragraph{Normalized Discounted Cumulative Gain (NDCG@$K$).} Since each test case has exactly one relevant (ground-truth) label, the ideal DCG@$K$ is always $1$ (achieved when the correct label is ranked first), and NDCG@$K$ reduces to:
\begin{equation}
\text{NDCG@}K = \frac{1}{N} \sum_{i=1}^{N}
\begin{cases}
\dfrac{1}{\log_2(\text{rank}_i + 1)} & \text{if } \text{rank}_i \leq K \\[6pt]
0 & \text{otherwise}
\end{cases}
\end{equation}
NDCG@$K$ provides a softer rank-sensitive score, rewarding correct predictions more when they appear near the top of the ranked list while penalizing lower-ranked matches less aggressively than MRR.

\subsection{Experimental Setup}
The experimental evaluation has been  conducted using an infrastructure composed of two virtual machines hosted on a Proxmox Virtual Environment (Proxmox VE) 9.1.7 server, running Linux kernel 6.17.13-2-pve on an x86\_64 architecture. The physical server is equipped with an AMD EPYC\texttrademark{} Genoa 9554P processor (64 cores/128 threads, 3.1\,GHz base frequency) and 192\,GB of RAM.

The first virtual machine, hosting the entire framework execution environment, runs Ubuntu 24.04.4 LTS (Linux kernel 6.17.0-35-generic) and is configured with 16 vCPUs (1 socket, 16 cores), 16\,GB of RAM, Docker 29.5.3, and LXC 5.21.5 LTS. The test environment is composed of four main components:
\begin{inparaenum}[(i)]
    \item the \textit{target} environment, representing the system under analysis (based on Ubuntu 24.04), which can be deployed either as a Docker container or as an LXC virtual machine. This environment is configured with the tools required to execute Atomic Red Team tests, including OpenSSH and PowerShell;
    \item the \textit{tracee} container, running \textit{tracee} v0.24.1, which leverages eBPF technology to monitor the activities performed within the \textit{target} environment, collecting logs and forwarding them in JSON format to \textit{fluentd};
    \item the \textit{fluentd} container, running \textit{fluentd} v1.19.3, responsible for managing the logs generated by \textit{tracee}, applying buffering techniques and log rotation mechanisms;
    \item the \textit{neo4j} container, running \textit{neo4j} v2026.05.0, used for storing and querying the generated provenance graph.
\end{inparaenum}

The second virtual machine is dedicated to the local execution of the LLMs employed during the analysis phases. It runs Ubuntu 24.04.4 LTS (Linux kernel 6.17.0-35-generic) and is configured with 16 vCPUs (1 socket, 16 cores), 32\,GB of RAM, LM Studio 0.4.16 (Build 2), NVIDIA driver 595.71.05, and CUDA 13.2. Model inference is performed entirely locally through LM Studio, leveraging two NVIDIA L40S GPUs with 48\,GB of memory each.

All models are served with the same context window of 32K tokens (Table 2). This value corresponds to the native context limit of foundation-sec-8b-reasoning, the most restrictive among the evaluated models, and was applied uniformly to all of them so that differences in classification quality reflect the models’ reasoning capabilities rather than the amount of input each one can process. The same budget is retained in the ablation study (Section~\ref{sec:ablation}), as it represents the deployment constraint under which any input representation must operate.

\subsection{Results}
\label{sec:results}

\begin{table}[ht]
\centering
\caption{Number of invalid JSON outputs for each model, with and without RAG.}
\small
\begin{tabular}{lcc}
\hline
& \textbf{Pure LLM} & \textbf{RAG} \\
\hline
deepseek-r1-distill-qwen-32b & 0 & 0 \\
foundation-sec-8b-reasoning & 16 & 20 \\
gemma-4-31b-it & 0 & 0 \\
gpt-oss-120b & 2 & 0 \\
gpt-oss-20b & 8 & 6 \\
llama-3.3-70b-instruct & 1 & 1 \\
qwen3.5-9b & 0 & 1 \\
\hline
\end{tabular}
\label{tab:invalid_json_llm_rag}
\end{table}

A key factor influencing \attack{} mapping quality is the amount of discriminative behavioral evidence contained in the input sequence. At the technique level, longer multi-step command traces often provide a richer causal context, making it easier to distinguish among broad \attack{} behaviors. By contrast, sub-technique classification is inherently more ambiguous, since sub-techniques capture finer-grained variations of a broader behavior and therefore require more specific contextual cues to disambiguate between command sequences that may be compatible with multiple related \attack{} classes~\cite{mitredesignphilosofy}. Recent work, directly from MITRE organization and the Center for Threat-Informed Defense~\cite{mitreAmbiguousTechniques} also highlights that many \attack{} techniques are ambiguous when considered in isolation and that surrounding context is necessary to infer the correct adversarial intent. 

\paragraph{\textbf{Label-identifier inconsistency.}}
During evaluation, we observed a recurring failure mode across all tested models: the model correctly identifies a technique by name but associates it with an incorrect technique ID, or conversely, produces the correct ID paired with a mismatched technique name. This inconsistency arises because the model relies on its parametric knowledge to map the observed behavior to a technique name and its corresponding identifier, but this internal mapping is not always reliable.
To mitigate this issue, we introduce a second prompt variant, termed \emph{taxonomy-grounded prompting}. In this configuration, the system prompt is augmented with the complete mapping of technique IDs to their canonical names, extracted directly from the MITRE ATT\&CK Enterprise knowledge base~\cite{mitrekb}. 
The injected mapping acts as a grounding signal that anchors the model's predictions to the official taxonomy.

Tables~\ref{tab:results_combined},~\ref{tab:results_rag}, summarize the evaluation results across the three configurations considered in this study. For paired comparisons, $\Delta$ indicates the absolute improvement in HR@$k$ of taxonomy-grounded prompting over base LLM prompting. 
% HR@$k$ measures the percentage of test cases where the correct label is within the top-$k$ predictions returned by the LLM (Appendix~\ref{sec:appendix_a}).
% In our experiments, w
We set $K=5$.

\paragraph{\textbf{LLM results.}}
Under base LLM prompting, performance varies substantially across models. Notably, foundation-sec-8b-reasoning -- an 8B model specialized on security-related pretraining data -- outperforms substantially larger general-purpose models such as llama-3.3-70b-instruct (Tech HR@5 $=49.2\%$ vs.\ $35.26\%$, a $+13.94$ point gap despite an almost $9\times$ parameter disadvantage) and deepseek-r1-distill-qwen-32b ($49.2\%$ vs.\ $29.97\%$), and is competitive with gpt-oss-20b. This suggests that domain-specialized pretraining can be an effective driver of \attack{} mapping quality. gpt-oss-120b remains the strongest model overall (Tech HR@5 $=60.58\%$, Subtech HR@5 $=33.33\%$), while deepseek-r1-distill-qwen-32b is the weakest.
Taxonomy-grounded prompting yields consistently higher scores across all models and settings, and gpt-oss-120b again outperforms the other models. On the exact sub-technique match level, it reaches HR@5 $=43.66\%$, the best result among all models. In the technique setting, gpt-oss-120b achieves the highest HR@5 ($64.64\%$), closely followed by gemma-4-31b-it ($63.69\%$), while deepseek-r1-distill-qwen-32b again records the lowest scores.

\paragraph{\textbf{RAG results.}} 

In this configuration, the system prompt is the same as in pure prompting (Appendix~\ref{sec:appendix_c}), while the user prompt contains the retrieved chunks, as shown in Appendix~\ref{sec:appendix_b}.

% The system prompt employed in this configuration is the same of pure prompting approach (see Appendix~\ref{sec:appendix_c}), while the user prompt contains the retrieved chunks from the RAG as reported in Appendix~\ref{sec:appendix_b}. 

By injecting relevant \attack{} knowledge at inference time, the model can compare the observed behavior against a narrower and better grounded set of candidate techniques, reducing semantic ambiguity and improving fine-grained sub-technique attribution. This effect is reflected in our results: on the exact sub-technique match level, gpt-oss-20b comes close to doubling its HR@5 relative to base prompting ($16.11\% \rightarrow 31.55\%$, $+95.85\%$ relative gain). Overall, gpt-oss-120b reaches the highest HR@5 at sub-techniques level ($46.26\%$), while in the technique setting it achieves the strongest overall result across all configurations (HR@5 $=66.57\%$) followed by gemma-4-31b-it (HR@$5 = 65.42$), while deepseek-r1-distill-qwen-32b remains the weakest performer.

\paragraph{\textbf{Comparison.}}
Comparing base LLM prompting directly against RAG, HR@5 improves consistently across all model-criteria combinations ($14$ out of $14$), a stronger and more uniform trend than what is observed for taxonomy-grounded prompting alone. At the technique level, HR@5 gains range from $+1.78$ (foundation-sec-8b-reasoning) to $+10.66$ \\(gemma-4-31b-it), with an average improvement of $\sim\!5.52$ points. At the exact sub-technique level, gains are larger and more variable, ranging from $+0.93$ (gemma-4-31b-it) to $+15.44$ (gpt-oss-20b), averaging $\sim\!8.03$ points, confirming that retrieval-grounded context is especially beneficial for disambiguating fine-grained behavioral variants. Notably, the largest technique-level gain and the largest sub-technique-level gain occur for different, mid-sized models, suggesting that RAG's benefit is not strictly tied to model scale but rather to how well a given model's parametric knowledge already covers the \attack{} taxonomy.

\begin{figure*}[ht]
  \centering
  \includegraphics[width=0.9\linewidth]{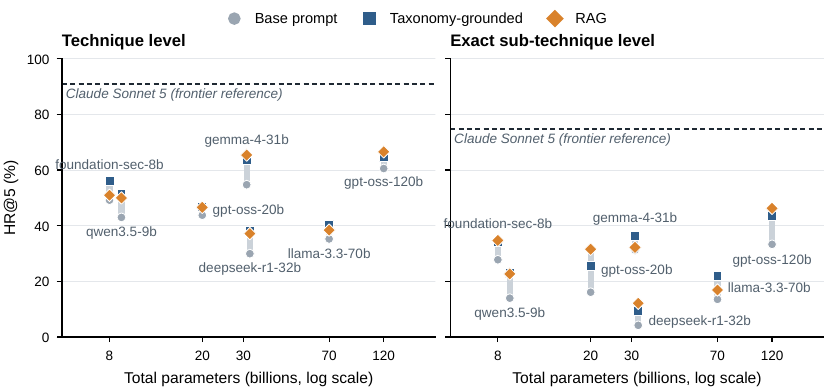}
  \caption{HR@5 of each locally deployed LLM under base, taxonomy-grounded and RAG prompting, at technique level (left) and exact sub-technique level (right). The grey segment spans the range across the three configurations. The dashed line marks the frontier-model reference (Claude Sonnet~5).}
  % \Description{Dot plot with one row per model and three markers per row (base, taxonomy-grounded, RAG). Grounded configurations lie to the right of base prompting for every model; all local models remain well below the frontier reference line.}
  \label{fig:hr5-config}
\end{figure*}

Figure~\ref{fig:hr5-config} summarizes the effect of the three prompting configurations. Both configurations improve over base prompting for every model and at both granularity levels (28 out of 28 paired comparisons). The spread between configurations is noticeably wider at the sub-technique level, where grounding matters most: gpt-oss-20b, for instance, moves from 16.11\% with base prompting to 31.55\% with RAG. 
%The best local configuration, gpt-oss-120b with RAG, reaches an HR@5 of 66.57\% at technique level and 46.26\% at sub-technique level, still 24.34 and 28.50 points below the frontier reference. Since the same pipeline and input representation are used in all cases, this gap is consistent with our interpretation that residual errors are driven mainly by model capacity rather than by the methodology.

\begin{figure}
  \centering
  \includegraphics[width=1\linewidth]{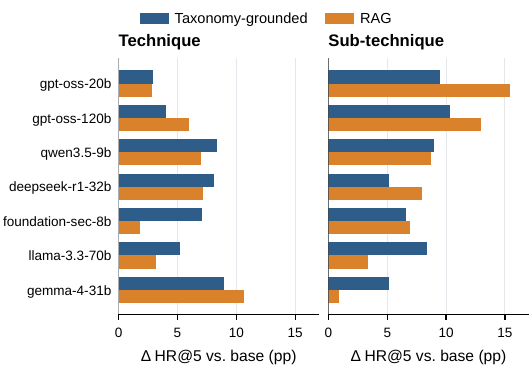}
  \caption{Absolute HR@5 gain (percentage points) of taxonomy-grounded prompting and RAG over base prompting, at technique (left) and exact sub-technique (right) level. Models are sorted by the sub-technique gain of RAG.}
  % \Description{Paired horizontal bar chart. All bars are positive. At technique level taxonomy-grounded bars are usually longer than RAG bars; at sub-technique level RAG bars are the longest for gpt-oss-20b and gpt-oss-120b.}
  \label{fig:gain}
\end{figure}

Figure~\ref{fig:gain} isolates the contribution of each grounding strategy. On average, taxonomy-grounded prompting improves HR@5 by $6.38$ points at technique level and $7.71$ points at sub-technique level, while RAG improves it by $5.52$ and $8.03$ points, respectively. The two strategies are complementary rather than interchangeable. Injecting the ID--name mapping directly addresses the label--identifier inconsistency and is the stronger option at technique level for five of the seven models. RAG, which provides technique descriptions in addition to the solely names, yields the largest gains at sub-technique level, up to $+15.44$ points for gpt-oss-20b and $+12.93$ points for gpt-oss-120b, where distinguishing closely related variants requires descriptive knowledge of each procedure. A notable exception is gemma-4-31b-it, whose sub-technique HR@5 barely changes with RAG ($+0.93$), suggesting that the benefit of retrieval depends on how well a model already encodes the \attack{} taxonomy.

\paragraph{\textbf{Successful vs. Failed Attacks.}}
An important practical question is whether \toolname{} remains useful when the underlying attack does not fully execute -- a common scenario in real deployments, where adversarial actions are frequently interrupted by defenses, misconfigurations, or missing prerequisites before completion. 

At the technique level, performance on failed attacks remains close to, and in some cases exceeds, performance on successful ones. For instance, under base prompting foundation-sec-8b reaches HR@5 $=49.67\%$ on successful attacks versus $48.75\%$ on failed ones. In general, results indicate that the coarse-grained adversarial intent underlying an \attack{} technique is often already legible from the partial command flow generated before an attack is interrupted, without requiring the full attack chain to unfold.

At the exact sub-technique level, the picture changes: the gap between successful and failed attacks widens substantially and is consistently in favor of successful attacks, and reaching over $21$ points for gemma-4-31b-it and, under RAG, over $29$ points for gpt-oss-120b (HR@5 $=60.36\%$ on successful attacks vs.\ $31.07\%$ on failed ones). This is consistent with the finer-grained nature of sub-technique classification discussed above: disambiguating between closely related behavioral variants requires more complete contextual evidence, which an interrupted execution, by definition, cannot fully provide.
Taken together, these results suggest that \toolname{} retains most of its value for technique-level triage and alerting, which is arguably the more actionable output for a SOC analyst reacting to an in-progress or blocked intrusion attempt, while its finer-grained sub-technique attribution capability is, as expected, more dependent on observing the complete adversarial procedure.
All these results and considerations allow us to answer positively to the first RQ: MITRE \attack{} techniques can be inferred with promising accuracy from kernel-level telemetry via LLM-based reasoning over reconstructed provenance graphs. 
% The frontier-model validation further confirms the soundness of the overall pipeline, indicating that the observed gap for local models stems from model capacity rather than methodological limitations.
\begin{figure*}
    \centering
    \includegraphics[width=1\linewidth]{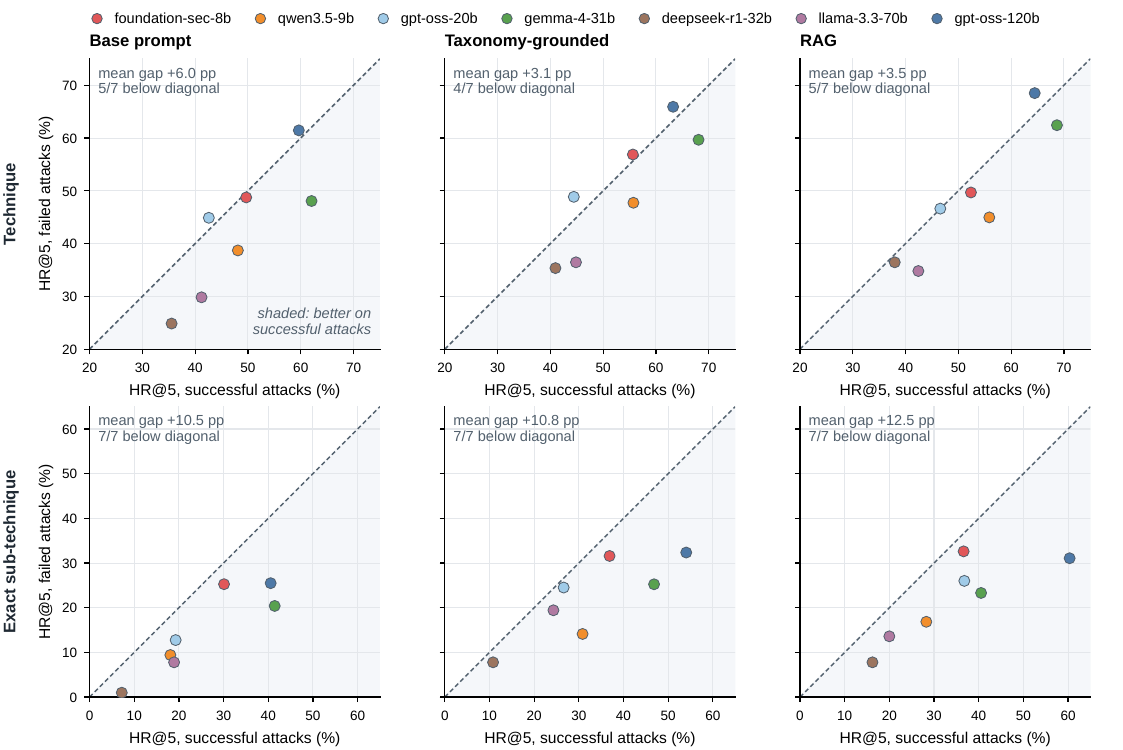}
    \caption{HR@5 on successfully executed attacks ($N=166$) versus failed attacks ($N=181$) for every model and configuration. Points below the diagonal perform better on successful attacks.}
    \label{fig:succ-fail}
\end{figure*}

Figure~\ref{fig:succ-fail} compares performance on attacks that completed with performance on attacks failed due to missing dependencies or permissions. At technique level, points cluster close to the diagonal, which represents equal performance on successful and failed attacks: the mean gap is $4.24$ points, and in 7 of the 21 model--configuration pairs the HR@5 on failed attacks is even higher than on successful ones. At sub-technique level, instead, all 21 pairs lie below the diagonal, with a mean gap of $11.26$ points that reaches $29.29$ points for gpt-oss-120b with RAG. This asymmetry confirms that the coarse adversarial intent is already legible from the partial command flow, whereas disambiguating between sibling sub-techniques requires observing the complete procedure.

\begin{table*}[ht]
\centering
\tiny
\caption{Mapping performance across LLMs: base vs. taxonomy-grounded prompt.}
\label{tab:results_combined}
\renewcommand{\arraystretch}{1.2}
\setlength{\tabcolsep}{3pt}
\begin{tabular}{l|l|c|ccc|ccc|c|ccc|ccc|c}
\toprule
\multirow{2}{*}{\textbf{Model}} & \multirow{2}{*}{\textbf{Level}} &
\multicolumn{7}{c|}{\textbf{Base Prompt}} &
\multicolumn{7}{c|}{\textbf{Taxonomy-Grounded Prompt}} &
\multirow{2}{*}{\shortstack{\textbf{$\Delta$ HR@5 }(\%)}} \\
& &
\textbf{All} & \multicolumn{3}{c|}{\textbf{Successful}} & \multicolumn{3}{c|}{\textbf{Failed}} &
\textbf{All} & \multicolumn{3}{c|}{\textbf{Successful}} & \multicolumn{3}{c|}{\textbf{Failed}} &
\\
& &
\textbf{HR@5} & \textbf{HR@5} & \textbf{MRR@5} & \textbf{NDCG@5} & \textbf{HR@5} & \textbf{MRR@5} & \textbf{NDCG@5} &
\textbf{HR@5} & \textbf{HR@5} & \textbf{MRR@5} & \textbf{NDCG@5} & \textbf{HR@5} & \textbf{MRR@5} & \textbf{NDCG@5} &
\\
\toprule
\multirow{2}{*}{\shortstack[l]{foundation-sec\\-8b-reasoning}}
& Tech &
49.20 & 49.67 & 0.342 & 0.381 & 48.75 & 0.332 & 0.371 & 56.27 & 55.63 & 0.368 & 0.416 & 56.88 & 0.395 & 0.438 & \textbf{+7.07} \\
& Subtech &
27.78 & 30.10 & 0.203 & 0.227 & 25.26 & 0.176 & 0.195 & 34.34 & 36.89 & 0.253 & 0.282 & 31.58 & 0.242 & 0.260 & \textbf{+6.56} \\
\midrule
\multirow{2}{*}{\shortstack[l]{qwen3.5-9b}}
& Tech &
43.01 & 48.09 & 0.350 & 0.382 & 38.71 & 0.298 & 0.321 & 51.40 & 55.73 & 0.400 & 0.439 & 47.74 & 0.348 & 0.380 & \textbf{+8.39} \\
& Subtech &
13.97 & 18.09 & 0.136 & 0.147 & 9.41 & 0.075 & 0.080 & 22.91 & 30.85 & 0.222 & 0.244 & 14.12 & 0.108 & 0.116 & \textbf{+8.94} \\
\midrule
\multirow{2}{*}{\shortstack[l]{gpt-oss-20b}}
& Tech &
43.79 & 42.59 & 0.305 & 0.335 & 44.89 & 0.338 & 0.366 & 46.75 & 44.44 & 0.318 & 0.350 & 48.86 & 0.382 & 0.409 & \textbf{+2.96} \\
& Subtech &
16.11 & 19.27 & 0.134 & 0.149 & 12.75 & 0.102 & 0.108 & 25.59 & 26.61 & 0.197 & 0.214 & 24.51 & 0.201 & 0.212 & \textbf{+9.48} \\
\midrule
\multirow{2}{*}{\shortstack[l]{gemma-4-31b-it}}
& Tech &
54.76 & 62.05 & 0.510 & 0.538 & 48.07 & 0.390 & 0.412 & 63.69 & 68.07 & 0.562 & 0.592 & 59.67 & 0.462 & 0.495 & \textbf{+8.93} \\
& Subtech &
31.31 & 41.44 & 0.335 & 0.355 & 20.39 & 0.183 & 0.188 & 36.45 & 46.85 & 0.382 & 0.404 & 25.24 & 0.225 & 0.232 & \textbf{+5.14} \\
\midrule
\multirow{2}{*}{\shortstack[l]{deepseek-r1-\\distill-qwen-32b}}
& Tech &
29.97 & 35.54 & 0.250 & 0.277 & 24.86 & 0.191 & 0.206 & 38.04 & 40.96 & 0.292 & 0.321 & 35.36 & 0.253 & 0.278 & \textbf{+8.07} \\
& Subtech &
4.21 & 7.21 & 0.053 & 0.058 & 0.97 & 0.005 & 0.006 & 9.35 & 10.81 & 0.081 & 0.088 & 7.77 & 0.063 & 0.067 & \textbf{+5.14} \\
\midrule
\multirow{2}{*}{\shortstack[l]{llama-3.3\\-70b-instruct}}
& Tech &
35.26 & 41.21 & 0.286 & 0.317 & 29.83 & 0.175 & 0.205 & 40.46 & 44.85 & 0.327 & 0.357 & 36.46 & 0.222 & 0.257 & \textbf{+5.20} \\
& Subtech &
13.55 & 18.92 & 0.129 & 0.144 & 7.77 & 0.053 & 0.059 & 21.96 & 24.32 & 0.151 & 0.174 & 19.42 & 0.121 & 0.139 & \textbf{+8.41} \\
\midrule
\multirow{2}{*}{\shortstack[l]{gpt-oss-120b}}
& Tech &
60.58 & 59.64 & 0.486 & 0.513 & 61.45 & 0.482 & 0.515 & 64.64 & 63.25 & 0.504 & 0.536 & 65.92 & 0.509 & 0.546 & \textbf{+4.06} \\
& Subtech &
33.33 & 40.54 & 0.344 & 0.360 & 25.49 & 0.195 & 0.210 & 43.66 & 54.05 & 0.445 & 0.470 & 32.35 & 0.265 & 0.280 & \textbf{+10.33} \\
% \bottomrule
% \multirow{2}{*}{\shortstack[l]{Claude-Sonnet-5}}
% & Tech &
% 60.58 & 59.64 & 0.486 & 0.513 & 61.45 & 0.482 & 0.515 & 64.64 & 63.25 & 0.504 & 0.536 & 65.92 & 0.509 & 0.546 & \textbf{+4.06} \\
% & Subtech &
% 33.33 & 40.54 & 0.344 & 0.360 & 25.49 & 0.195 & 0.210 & 43.66 & 54.05 & 0.445 & 0.470 & 32.35 & 0.265 & 0.280 & \textbf{+10.33} \\
\bottomrule
\end{tabular}
\end{table*}

\begin{table}[ht]
\centering
\tiny
\caption{Mapping performance across LLMs with RAG.}
\label{tab:results_rag}
\renewcommand{\arraystretch}{1.15}
\setlength{\tabcolsep}{1.2pt}
\begin{tabular}{l|l|c|ccc|ccc}
\toprule
\multirow{2}{*}{\textbf{Model}} & \multirow{2}{*}{\textbf{Level}} & \multicolumn{1}{c|}{\textbf{All Attacks}} & \multicolumn{3}{c|}{\textbf{Successful Attacks}} & \multicolumn{3}{c}{\textbf{Failed Attacks}} \\
 & & HR@5 (\%) & HR@5 (\%) & MRR@5 & NDCG@5 & HR@5 (\%) & MRR@5 & NDCG@5 \\
\toprule
\multirow{2}{*}{\shortstack[l]{foundation-sec\\-8b-reasoning}}
& Tech &
50.98 & 52.38 & 0.370 & 0.408 & 49.69 & 0.333 & 0.374 \\
& Subtech &
34.72 & 36.63 & 0.290 & 0.309 & 32.61 & 0.246 & 0.266 \\
\midrule
\multirow{2}{*}{\shortstack[l]{qwen3.5-9b}}
& Tech &
50.00 & 55.86 & 0.434 & 0.465 & 44.97 & 0.322 & 0.353 \\
& Subtech &
22.68 & 28.28 & 0.210 & 0.228 & 16.84 & 0.120 & 0.132 \\
\midrule
\multirow{2}{*}{\shortstack[l]{gpt-oss-20b}}
& Tech &
46.61 & 46.58 & 0.367 & 0.392 & 46.63 & 0.367 & 0.392 \\
& Subtech &
31.55 & 36.79 & 0.299 & 0.316 & 26.00 & 0.218 & 0.229 \\
\midrule
\multirow{2}{*}{\shortstack[l]{gemma-4-31b-it}}
& Tech &
65.42 & 68.67 & 0.550 & 0.585 & 62.43 & 0.488 & 0.522 \\
& Subtech &
32.24 & 40.54 & 0.312 & 0.336 & 23.30 & 0.179 & 0.193 \\
\midrule
\multirow{2}{*}{\shortstack[l]{deepseek-r1-\\distill-qwen-32b}}
& Tech &
37.18 & 37.95 & 0.269 & 0.297 & 36.46 & 0.269 & 0.293 \\
& Subtech &
12.15 & 16.22 & 0.129 & 0.137 & 7.77 & 0.045 & 0.054 \\
\midrule
\multirow{2}{*}{\shortstack[l]{llama-3.3-\\70b-instruct}}
& Tech &
38.44 & 42.42 & 0.314 & 0.341 & 34.81 & 0.193 & 0.231 \\
& Subtech &
16.90 & 20.00 & 0.137 & 0.152 & 13.59 & 0.073 & 0.088 \\
\midrule
\multirow{2}{*}{\shortstack[l]{gpt-oss-120b}}
& Tech &
66.57 & 64.46 & 0.505 & 0.540 & 68.51 & 0.515 & 0.558 \\
& Subtech &
46.26 & 60.36 & 0.489 & 0.518 & 31.07 & 0.267 & 0.278 \\
% \bottomrule
% \multirow{2}{*}{\shortstack[l]{Claude-Sonnet-5}}
% & Tech &
% 66.57 & 64.46 & 0.505 & 0.540 & 68.51 & 0.515 & 0.558 \\
% & Subtech &
% 46.26 & 60.36 & 0.489 & 0.518 & 31.07 & 0.267 & 0.278 \\
\bottomrule
\end{tabular}
\end{table}

\begin{figure*}[t]
  \centering
  \includegraphics[width=\linewidth]{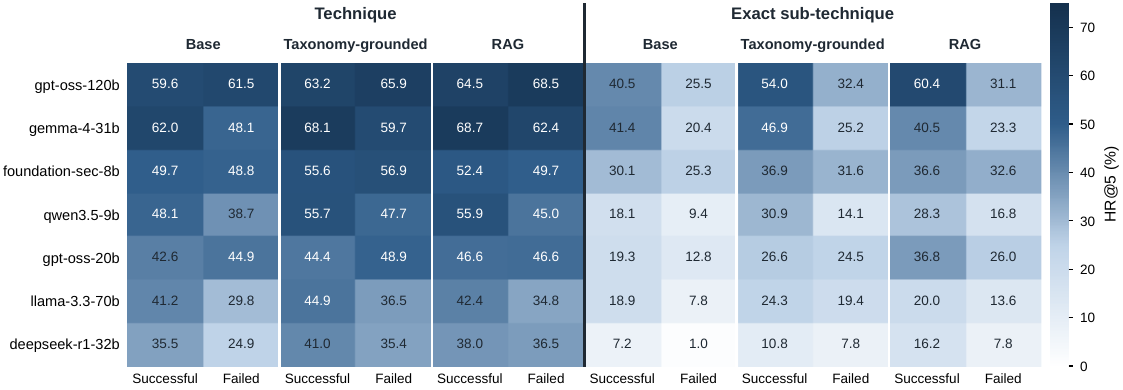}
  \caption{HR@5 (\%) of all models under the three prompting configurations, at technique (left) and exact sub-technique (right) level, on successful and failed attacks. Models are sorted by technique-level HR@5 with RAG.}
  % \Description{Annotated heatmap with seven rows (models) and eighteen columns. The left half (technique level) is mostly dark; the right half (sub-technique level) is lighter, especially in the failed-attack columns.}
  \label{fig:heatmap}
\end{figure*}
 
Figure~\ref{fig:heatmap} provides an overview of all HR@5 results. Two patterns stand out. First, model ranking is largely stable across configurations and subsets: gpt-oss-120b ranks first or second in all 12 columns, while deepseek-r1-distill-qwen-32b is the weakest model in 11 of them. Second, the hardest setting is exact sub-technique identification on failed attacks, where only gpt-oss-120b and foundation-sec-8b exceed $30\%$ HR@5 in any configuration.

\paragraph{\textbf{RQ2: The cost of local inference.}}
\toolname{} targets locally deployed open-weight models, since telemetry cannot leave the defender's trust boundary (Section~\ref{sec:confidentiality}). To quantify what this constraint costs, we run the same pipeline with a frontier cloud model, Claude-Sonnet-5, as an upper-bound reference. This experiment is compatible with our confidentiality requirement: Atomic Red Team is a public emulation library and our telemetry was generated in an isolated laboratory, so it contains no production secrets or personal data. The frontier model is not a deployment option for \toolname{}.

Sonnet-5 reaches an HR@5 of $90.91\%$ at the technique level and $74.76\%$ at the exact sub-technique level. This is $+24.34$ and $+28.50$ percentage points above the best local configuration (gpt-oss-120b with RAG). Retrieval narrows this gap but does not close it: for gpt-oss-120b, RAG recovers about $20\%$ of the technique-level gap and $31\%$ of the sub-technique-level gap with respect to base prompting. RAG yields no measurable gain for Sonnet-5, which suggests that the knowledge it injects is already encoded in the frontier model's parameters.

These results suggest that the pipeline produces representations from which \attack{} techniques can be inferred accurately when sufficient model capacity is available, and that the residual gap of local models is mainly a matter of capacity. At the same time, the gap quantifies the price of confidentiality under current open-weight models, which we expect to shrink as these models improve. 

\paragraph{\textbf{Discussion.}}

\begin{figure*}
  \centering
  \includegraphics[width=\linewidth]{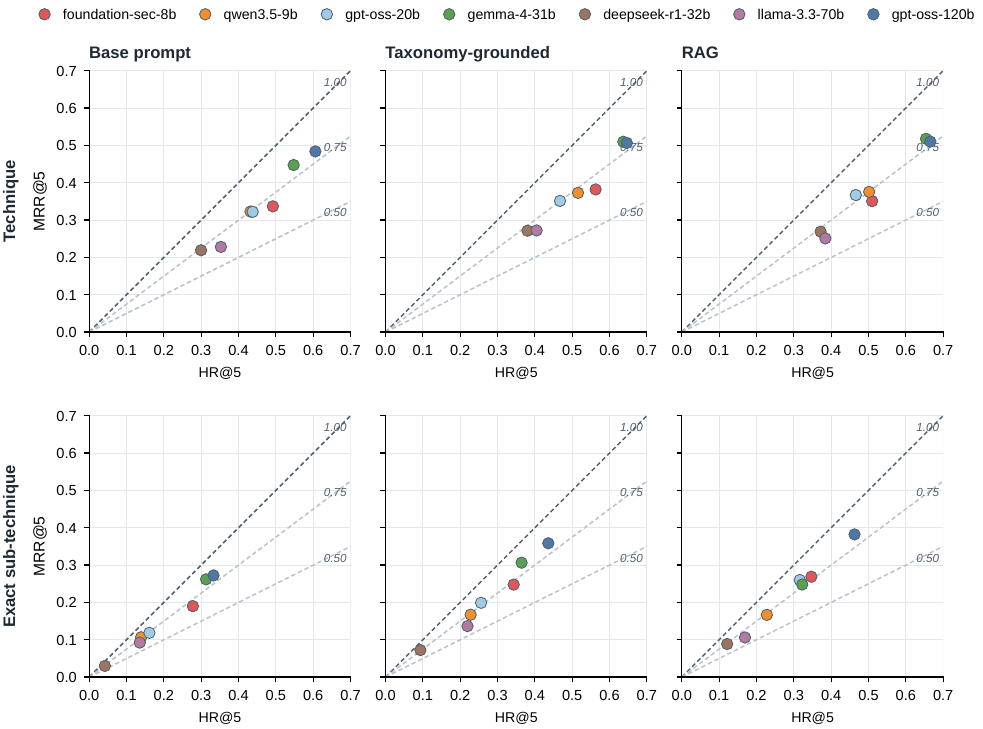}
  \caption{MRR@5 versus HR@5 for every model and configuration. Dashed iso-lines indicate the ratio MRR@5/HR@5, i.e., the mean reciprocal rank of the correct label among the cases in which it appears in the top-5 (1 = always ranked first). Values for all attacks are obtained by weighting successful and failed attacks by their sample size.}
  % \Description{Scatter plot of MRR against HR with diagonal iso-lines. All points lie between the 0.5 and 1.0 iso-lines, most of them close to the 0.75 line, regardless of configuration.}
  \label{fig:rank-quality}
\end{figure*}
 
HR@5 alone does not reveal \emph{where} the correct technique is placed within the candidate list. Figure~\ref{fig:rank-quality} relates MRR@5 to HR@5: the ratio between the two is equivalent to the mean reciprocal rank of the correct label among the hits. Across all models and configurations this ratio averages $0.74$ (range $0.62-0.84$), indicating that, when the correct technique is retrieved, it is typically ranked first or second. The ratio is essentially unchanged across base, taxonomy-grounded and RAG prompting ($0.73-0.76$ on average), showing that grounding increases the number of correct candidates rather than reordering them. The lowest ratios are observed for llama-3.3-70b-instruct and foundation-sec-8b-reasoning (below $0.69$ in most settings): for these models the correct label more often appears in the lower positions of the list, so their HR@5 overstates their usefulness when only the top-ranked candidate is consumed, e.g., by an automated alerting pipeline.

\paragraph{\textbf{Comparison with LADE.}}
Among prior work, LADE~\cite{lade} is the closest to our approach, as it similarly applies LLM-based reasoning directly to command-level evidence rather than to CTI reports. However, LADE is evaluated on a substantially smaller and less realistic scale: only $35$ attack sequences, each averaging approximately $990$ code snippets with just $4.8$ non-empty lines per snippet ($\sim\!4.7K$ lines of logs each). In contrast, our evaluation spans $347$ Linux Atomic Red Team test cases, each derived from a full kernel-level telemetry trace averaging $\sim\!62K$ lines of raw eBPF events per scenario -- more than one order of magnitude larger per scenario than LADE's snippet sequences. This makes our evaluation setting closer to a realistic operational deployment.

\subsection{Ablation Study}
\label{sec:ablation}
We performed two ablations to quantify the contribution of the graph-based representation in \toolname{}. The ablation study was conducted on gpt-oss-120b with RAG, the best-performing LLM identified in the main experiments. 

\paragraph{\textbf{Raw logs vs.\ graph-based representation.}}
We assess whether the graph construction step introduced in Section~\ref{sec:methodology} is actually necessary, or whether an LLM can perform comparably well when given the raw, chronologically ordered kernel-level event stream. 

Directly prompting the LLM with raw logs led to a clear degradation in ranking quality with respect to the graph-based representation. Raw eBPF output is significantly more verbose per unit of information, since it repeats process/user metadata and low-level syscall arguments across many redundant events; this inflates the prompt and, for longer attack sequences, risks exceeding the model's effective context window or forcing truncation, which directly harms recall of relevant behavior. Second, and more importantly, raw logs do not make causal relationships between events explicit -- the LLM must itself infer which process spawned which execution, and how a sequence of syscalls composes into a coherent adversarial procedure.
Raw logs exceed the context window, and therefore produce an empty output, in $56.73\%$ of cases at the technique level and $53.1\%$ at the sub-technique level. Among the remaining cases, HR@5 is $37.88\%$ at the technique level and $19.57\%$ at the sub-technique level (exact match), i.e., only $16.4\%$ and $9.2\%$ of all test cases receive a correct mapping.
% Raw logs achieve $19.57\%$ HR@5, 0.214 MRR@5, and 0.227 NDCG@5, with empty outputs in $53.1\%$ cases at the sub-technique level (exact match), while at the technique level raw logs achieve $37.88\%$ HR@5, 0.298 MRR@5, and 0.385 NDCG@5, with empty outputs for token-limit exceeding, in $56.73\%$ of cases. 
This confirms that unstructured telemetry is difficult to map reliably to ATT\&CK techniques and, more importantly, that the raw-log baseline can fail when the input approaches the context limit of the local inference settings.

\paragraph{\textbf{Full Provenance Graph vs.\ Execution Graph.}}
We compared the full PG against the reduced Execution Graph used in the main experiments under a fixed context budget of 32K tokens. The proposed full PG correlation achieves $46.81\%$ HR@5 at sub-techniques level with exact match, while at technique level $69.27\%$ HR@5 when the mapping occurs without reaching context-window limit. 
Although the full PG achieves higher raw HR@5 whenever it fits within the context window, this advantage is undermined by its practical unreliability: with nearly one in two inputs producing an empty output (up to $49.76\%$ at the technique level and $45.93\%$ at the sub-technique level), the full PG effectively fails to return \emph{any} mapping for roughly half of the test cases. In an operational setting, a technique-level prediction that is only slightly less accurate but consistently available is far more valuable than a marginally better prediction that is missing half of the time. The compact execution-graph representation, by contrast, never exceeds the context budget and therefore always yields a usable candidate list, making it the more practical choice for deployment under bounded local inference resources, even at the cost of a modest drop in per-case accuracy.

\begin{figure*}
  \centering
  \includegraphics[width=\linewidth]{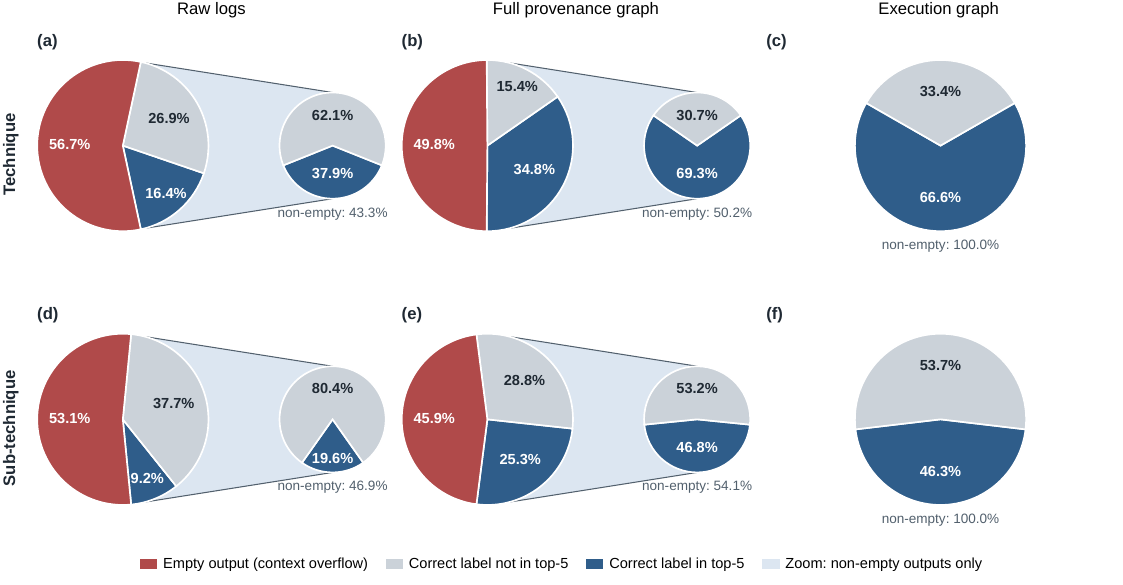}
   \caption{Ablation outcomes for raw logs, full provenance graph and execution graph (gpt-oss-120b with RAG, 32K-token context) at technique (a–c) and sub-technique (d–f) level. Each large pie splits all test cases into empty outputs caused by inputs exceeding the context window (red), non-empty outputs without the correct label in the top-5 (grey), and with the correct label in the top-5 (blue, i.e. HR@5 over all cases). The small pie expands the non-empty outputs to 100\%; its blue share is the HR@5 among non-empty outputs. The execution graph never overflows, so a single pie is shown; it yields 1.9× (technique) and 1.8× (sub-technique) the correct mappings of the full provenance graph.}
  
  % \Description{Stacked bar chart with three bars per panel. For raw logs and the full provenance graph about half of each bar is hatched (no output); the command graph bar has no hatched part and the largest blue segment.}
  \label{fig:ablation}
\end{figure*}

% The choice of command-graph representation has a substantial effect on classification quality. Raw, unstructured telemetry performs markedly worse (HR@5 $=19.57\%$ sub-technique / $37.88\%$ technique) and frequently produces empty outputs due to context-window truncation. Full provenance graphs improve accuracy over raw logs but remain token-heavy, causing empty outputs in up to $49.76\%$ of cases; the compact, weighted command-graph representation used in our main experiments avoids this failure mode entirely ($0.0\%$ empty outputs). Figure~\ref{fig:ablation} summarizes both ablations. 
\paragraph{\textbf{RQ3: The effectiveness of graph-based representation.}}
Graph-based representation is essential under a bounded context: raw, unstructured telemetry performs \\
markedly worse (HR@5 = $19.57\%$ sub-technique / $37.88\%$ technique) and exceeds the context window in more than half of the cases, making flat event streams impractical for local inference.
Among graph representations, the choice of abstraction level matters equally. Full provenance graphs improve accuracy over raw logs but remain token-heavy, causing empty outputs in up to $49.76\%$ of cases. The compact, weighted execution-graph representation used in our main experiments resolves this failure mode entirely ($0.0\%$ empty outputs), trading a modest drop in per-case accuracy for consistent, deployment-ready output under bounded inference resources. Figure~\ref{fig:ablation} summarizes both ablations.

%The full PG attains the highest HR@5 ($69.27\%$ at technique level and $46.81\%$ at sub-technique level), but it produces no output in $49.76\%$ and $45.93\%$ of the cases, respectively; raw logs are both less accurate and similarly unreliable. The command graph, with an HR@5 of 64.64\% and 43.66\%, is the only representation that returns a candidate list for every test case.

\section{Related Work}
\label{sec:related-work}

\begin{table}[t]
\centering
\caption{Comparison of Trace2ATT\&CK with representative approaches.}
\label{tab:related-work-comparison}
\tiny
\setlength{\tabcolsep}{2pt}
\renewcommand{\arraystretch}{1.15}
\begin{tabular}{lccccc}
\toprule
\textbf{Approach} &
\shortstack{\textbf{Runtime}\\\textbf{telemetry}} &
\shortstack{\textbf{Provenance}\\\textbf{context}} &
\shortstack{\textbf{Session-level}\\\textbf{analysis}} &
\shortstack{\textbf{LLM}\\\textbf{reasoning}} &
\shortstack{\textbf{Ranked}\\\textbf{techniques}} \\
\midrule

% CamQuery~\cite{pasquier2018runtime}
% & \cmark & \cmark & \xmark & \xmark & \xmark & \xmark \\

MGDA~\cite{CUI2026111806}
& \cmark & \cmark & \xmark & \xmark & \xmark \\

% KAIROS~\cite{cheng2024kairos}
% & \cmark & \cmark & \xmark & \xmark & \xmark & \xmark \\

CTINexus~\cite{cheng2025ctinexus}
& \xmark & \xmark & \xmark & \cmark & \xmark \\

SynthCTI~\cite{ruiz2026synt}
& \xmark & \xmark & \xmark & \cmark & \xmark \\

% PROVCON~\cite{yusof2025observations}
% & \cmark & \cmark & \xmark & \xmark & \xmark & \xmark \\

LADE~\cite{lade}
& \cmark & \xmark & \cmark & \cmark & \cmark \\

RAM~\cite{wudali2025rule}
& \xmark & \xmark & \xmark & \cmark & \cmark \\

\midrule
\textbf{Trace2ATT\&CK}
& \textbf{\cmark}
& \textbf{\cmark}
& \textbf{\cmark}
& \textbf{\cmark}
& \textbf{\cmark} \\
\bottomrule
\end{tabular}

\vspace{1mm}
\parbox{0.97\columnwidth}{\footnotesize
\textbf{\textit{Runtime telemetry}} indicates that the approach operates on execution traces rather than CTI reports or analyst-authored rules.
\textbf{\textit{Provenance context}} denotes explicit causal dependencies among system entities.
\textbf{\textit{Session-level analysis}} refers to reasoning over a bounded attacker execution session.
\textbf{\textit{Ranked techniques}} indicates that the approach returns multiple ATT\&CK techniques ordered by confidence.}
\end{table}

The automatic identification of MITRE \attack{} techniques from security data is an area of growing interest for both research and industrial applications. This is due to the fact that it supports activities such as threat hunting, incident response, and the assessment of detection capability coverage.

Various approaches have been developed to address this issue, with the utilization of CTI reports serving as the primary information source. In particular, LLMs have recently been employed for the automatic extraction of entities, relationships, and concepts from unstructured information, facilitating the construction of \textit{knowledge graphs} and the organization of knowledge~\cite{liu2024llmtikg, CTI-Thinker}. 

% \paragraph{\textbf{Provenance-Based Investigation}}
% System provenance represents causal dependencies among processes, files, and network endpoints as directed graphs, supporting information-flow tracing and attack reconstruction. Han et al.~\cite{han2018provenance} formulate provenance-based intrusion detection as graph-based anomaly detection, arguing that causal dependencies and long-range correlations help identify anomalous attack subgraphs. They also highlight the challenges posed by large, dynamic, and attributed provenance graphs.
% Pasquier et al.~\cite{pasquier2018runtime} introduce CamQuery, a framework for in-kernel and user-space analysis of live whole-system provenance streams.
% By integrating capture and graph processing at runtime, CamQuery supports low-latency security applications, including information-flow control, anomaly detection, and post-hoc forensic analysis. 
% These works establish provenance as a foundation for detecting, tracing, and explaining suspicious system activity.
% More recent systems build on these principles. MGDA~\cite{CUI2026111806} learns structural and contextual representations to identify anomalous provenance entities, then applies rule-based matching to reconstruct TTP-oriented scenarios. Similarly, KAIROS~\cite{cheng2024kairos} analyzes whole-system provenance to detect anomalous behavior and generate compact attack traces. These approaches operate on broad audit streams and jointly address attack localization and anomaly detection.

% \paragraph{\textbf{\attack{} Mapping from CTI and Detection Content}}
MGDA~\cite{CUI2026111806} learns structural and contextual representations to identify anomalous provenance entities, then applies rule-based matching to reconstruct TTP-oriented scenarios. CTINexus~\cite{cheng2025ctinexus} leverages in-context learning to extract entity--relation triplets from CTI reports, resolve semantically equivalent entities, and infer missing relations. To mitigate the scarcity of labeled data for CTI-to-\attack{} classification, SynthCTI~\cite{ruiz2026synt} generates semantically guided synthetic samples that improve the coverage of underrepresented techniques. Likewise, LLM-based threat-hunting frameworks integrate semantic reasoning, threat intelligence, and graph-based correlation to produce investigation outputs aligned with the \attack{} framework~\cite{manikandan2026agentic}.
%PROVCON~\cite{yusof2025observations} bridges CTI and system-level evidence by extracting attack primitives from reports, reproducing campaigns in a cyber range, and deriving provenance graphs from the resulting telemetry. It therefore relies on \emph{a priori} campaign knowledge to generate representative evidence. Our work instead starts from observed runtime activity and infers \attack{} candidates without requiring a campaign-specific CTI description to drive execution.

Recent work has also applied LLMs to command-level \attack{} analysis. LADE~\cite{lade} analyzes chronologically ordered command and script snippets to detect APT activity, localize malicious snippets, and rank corresponding \attack{} techniques. It combines rubric-based prompting, ATT\&CK knowledge, and summary propagation to preserve context across long sequences. Unlike LADE, which jointly performs detection, localization, and mapping over logged code snippets, \toolname{} assumes a pre-delimited attack interval and derives its mapping input from a provenance-grounded reconstruction of shell and descendant-process execution. This provides explicit causal relations from kernel-level telemetry rather than relying only on chronological command context.

Related approaches also map detection artifacts to \attack{}. RAM~\cite{wudali2025rule} uses an LLM pipeline that combines SIEM-rule descriptions, \attack{} retrieval, candidate generation, and relevance filtering. SIEM and detection-coverage platforms likewise associate telemetry with \attack{} through curated rules and correlation logic. However, such rules encode analyst-defined detection hypotheses, whereas runtime events acquire meaning only through their command, process, and temporal context. Consequently, mapping provenance derived execution flows differs from mapping rule text or detection metadata.

% Overall, prior work either detects and explains suspicious provenance activity, extracts \attack{} knowledge from high-level descriptions, or maps curated detection content. \toolname{} complements these directions by supporting provenance-grounded \attack{} mapping from observed session-level runtime behavior.

% %%%%%%%%%%%%%%%%%%%%%%%%%
% Unlike prior work that applies LLMs to narrative CTI, heterogeneous SOC
% telemetry, or post-detection graph summarization, \toolname{} applies
% LLM-based reasoning to compact, provenance-aware command graphs derived
% from eBPF telemetry. This formulation enables the empirical comparison
% of pure LLM prompting and retrieval-grounded reasoning for ranked
% \attack{} technique and sub-technique attribution.
% %%%%%%%%%%%%%%%%%%%%%%%%%

%Overall, prior work has focused on detecting and explaining suspicious provenance activity, extracting \attack{} knowledge from high-level descriptions, or mapping curated detection content. 
%In contrast, \toolname{} performs provenance-grounded \attack{} mapping from observed session-level runtime behavior. 
%Unlike prior work that applies LLMs to narrative CTI, heterogeneous SOC telemetry, or post-detection graph summarization, \toolname{} applies LLM-based reasoning to compact command graphs derived from eBPF telemetry, enabling an empirical comparison between pure prompting and retrieval-grounded reasoning for ranked \attack{} technique and sub-technique attribution.

\section{Conclusion}
\label{sec:conclusion}
This paper presented \toolname{}, an end-to-end framework that maps kernel-level telemetry to MITRE \attack{} techniques without moving that telemetry outside the defender's infrastructure. \toolname{} collects system events via eBPF and reconstructs the attacker's activity as a process-centric provenance graph rooted at the attacker shell. It then condenses this graph into a compact, weighted execution graph and classifies it with locally deployed open-weight LLMs, either with pure prompting or with RAG grounded in the \attack{} knowledge base.
Our evaluation on 347 Linux Atomic Red Team scenarios with seven open-weight LLMs answers the three research questions as follows.

\textbf{RQ1.} \attack{} techniques can be inferred from kernel-level telemetry with local models. The best configuration (gpt-oss-120b with RAG) reaches an HR@5 of 66.57\% at the technique level. RAG improves over base prompting in every model and setting, by up to $+15.44$ points at the sub-technique level. A security-specialized 8B model rivals much larger general-purpose ones, suggesting that domain pretraining can matter more than scale. Technique-level accuracy is largely preserved even when the attack is interrupted, whereas sub-technique attribution benefits from observing the complete procedure.

\textbf{RQ2.} A frontier cloud model, used only as an upper-bound reference on public laboratory data, outperforms the best local configuration by 24.34 (technique) and 28.50 (sub-technique) percentage points. This gap quantifies the current cost of keeping inference local. Retrieval recovers part of it, and we expect it to shrink further as open-weight models improve.

\textbf{RQ3.} The representation matters. Raw telemetry yields the lowest accuracy and frequently exceeds the context window. The execution graph always fits the context budget and yields the most correct mappings overall (1.9× the full provenance graph at technique level).

\paragraph{\textbf{Limitations and future work.}}
Our evaluation relies on Atomic Red Team, whose tests have a single ground-truth label. In addition, the process-centric community follows process lineage, so activity that escapes it, for example a payload later executed by \texttt{cron} or \texttt{systemd}, is not captured. Future work will address multi-stage campaigns with multiple ground-truth techniques and information flows across processes.

Data contamination represents a further limitation. As discussed in Section~\ref{sec:dataset}, although several factors reduce this risk, we cannot rule out memorization for scenarios whose commands have remained stable across releases. Future work could address this by evaluating on newly crafted attack scenarios not previously seen during model pre-training.

\section*{Acknowledgment}
This work has been partially supported within the technology transfer activities of the Second-Level Master’s Program in Electric Mobility and Circular Economy framework, funded by the Italian MUR under the “Territorial Pacts for Higher Education for Enterprises” (CUP H52C23000090001).
The work of M. Lupinacci and L. Arena was supported by Agenzia per la cybersicurezza nazionale under the program for the promotion of XLI cycle PhD research in cybersecurity – (CUP H23C25000360005, CUP H23C26000400005). The views expressed are those of the authors and do not represent the funding institution.
% To print the credit authorship contribution details
\printcredits

%% Loading bibliography style file
%\bibliographystyle{model1-num-names}
\bibliographystyle{cas-model2-names}

% Loading bibliography database
\bibliography{cas-refs}

\appendix

\section{RAG Prompt}
\label{sec:appendix_b}
\begin{tcolorbox}[
    colback=blue!5!white,
    colframe=blue!75!black,
    title=Example of user input prompt
]
\scriptsize
Context: \{retrieved chunks from RAG\}\\
Question: \{Execution Graph: \{dot\_graph representation\} \}\\
\end{tcolorbox}

\section{System Prompt}
\label{sec:appendix_c}
\begin{tcolorbox}[
    colback=blue!5!white,
    colframe=blue!75!black,
    title=LLM System Prompt
]
\scriptsize
You are a cyber threat intelligence analyst analyzing an attack sequence received as a DOT file. The DOT file represents the FULL attack sequence as a complete execution graph with interconnected nodes and edges. \\

\textbf{Your Task \\}
1. Read and analyze the ENTIRE graph comprehensively - examine ALL nodes, ALL edges, and the COMPLETE flow from start to finish \\
2. Map the WHOLE attacker sequence (not isolated commands) to MITRE ATT\&CK techniques \\
3. For each mapping, consider how commands relate to each other across the entire workflow \\
4. Select the top 5 most appropriate techniques.

\textbf{Critical Requirements for Each Technique: \\}
- MUST identify the complete TACTIC \\
- MUST identify the complete TECHNIQUE with sub-technique in format: 'Technique Name (ID)' where ID includes sub-technique (e.g., "T1003.007" not just "T1003") \\
- sub-techniques are REQUIRED: If a technique has sub-techniques and the attack shows sub-technique-level behavior, you MUST map to the sub-technique (e.g., T1003.007 "Proc Filesystem" not just T1003 "OS Credential Dumping") \\
- confidence: a (0,1) value reflecting how well the ENTIRE flow supports this mapping \\
- rationale: Explain WHY this mapping fits based on the HOLISTIC attack pattern across the entire graph flow, NOT based on a single isolated command

\textbf{Rationale Writing Guidelines: \\}
- BAD: "Mapping chosen due to presence of 'sh' command" \\
- GOOD: "Mapping chosen because the entire flow shows process memory dumping through sequential commands: target process launch → sh-based memory extraction → credential file creation, indicating a complete OS credential dumping workflow" \\
- BAD: "Found T1003 because of dump command" \\ 
- GOOD: "The complete attack sequence demonstrates credential dumping through the entire workflow: initial process targeting, memory extraction via shell, and subsequent credential file manipulation, matching T1003.007 Proc Filesystem sub-technique"

\textbf{Output Rules: \\}
Each element of the output is a JSON entry containing: \\
- "tactic": the ATT\&CK tactic name \\
- "technique": the ATT\&CK technique with sub-technique name and ID in format 'Technique Name (ID)' (e.g., "OS Credential Dumping: Proc Filesystem (T1003.007)") \\
- "confidence": a (0,1) value \\
- "rationale": explanation based on the COMPLETE attack flow, not isolated commands
Return ONLY a JSON array containing the 5 candidates ordered by confidence descending. NO text, explanations, or markdown outside the JSON array. 
\end{tcolorbox}

% Biography
%\bio{}
% Here goes the biography details.
%\endbio

%\bio{pic1}
% Here goes the biography details.
%\endbio

\end{document}